\documentclass[aps,prc,twocolumn,nofootinbib,amsmath,superscriptaddress,floatfix]{revtex4-2}
\usepackage{graphicx,epsfig,longtable}
\usepackage{mathptmx}
\usepackage[light,first]{draftcopy} 
\usepackage{epstopdf}
\usepackage[pdftex]{hyperref}
\usepackage{wasysym}
\usepackage{multirow}
\usepackage{eqnarray,amsmath,amsbsy}
\usepackage{booktabs}
\usepackage{appendix}
\usepackage{xcolor}

\newcommand{\beqy}{\begin{eqnarray}}
\newcommand{\eeqy}{\end{eqnarray}}
\newcommand{\bmlet}{\begin{subequations}}
\newcommand{\emlet}{\end{subequations}}
\newcounter{saveeqn}

\def\gsimeq{\,\,\raise0.14em\hbox{$>$}\kern-0.76em\lower0.28em\hbox
{$\sim$}\,\,}
\def\lsimeq{\,\,\raise0.14em\hbox{$<$}\kern-0.76em\lower0.28em\hbox
{$\sim$}\,\,}

\begin{document}


\title{Statistical and non-statistical $\gamma$-decay properties of $^{64}$Zn}

\author{A.~C.~Larsen}
\email{a.c.larsen@fys.uio.no}
\affiliation{Department of Physics, University of Oslo, N-0316 Oslo, Norway}
\affiliation{Norwegian Nuclear Research Centre, Oslo, Norway}

\author{M.~Guttormsen}
\email{magne.guttormsen@fys.uio.no}
\affiliation{Department of Physics, University of Oslo, N-0316 Oslo, Norway}
\affiliation{Norwegian Nuclear Research Centre, Oslo, Norway}

\author{T.~K.~Eriksen}
\affiliation{Department of Physics, University of Oslo, N-0316 Oslo, Norway}
\affiliation{Norwegian Nuclear Research Centre, Oslo, Norway}
\affiliation{Department of Radiation Protection and Physics, Division NUK Kjeller,
Institute for Energy Technology, 2007 Kjeller, Norway}

\author{G.~M.~Tveten}
\affiliation{Department of Physics, University of Oslo, N-0316 Oslo, Norway}
\affiliation{Expert Analytics AS, N-0179 Oslo, Norway}

\author{H.~Utsunomiya}
\affiliation{Department of Physics, Konan University, Okamoto 8-9-1, Higashinada, Kobe 658-8501, Japan}
\affiliation{Shanghai Advanced Research Institute, Chinese Academy of Sciences, Shanghai 201210, China}

\author{J.~K.~Dahl}
\affiliation{Department of Physics, University of Oslo, N-0316 Oslo, Norway}
\affiliation{Norwegian Nuclear Research Centre, Oslo, Norway}

\author{N.~Shimizu}
\affiliation{Center for Computational Sciences, University of Tsukuba, Tsukuba, Ibaraki 305-8577, Japan}
\affiliation{Faculty of Pure and Applied Sciences, University of Tsukuba, Tsukuba, Ibaraki 305-8577, Japan}
\affiliation{Center for Nuclear Study, The University of Tokyo, 7-3-1, Hongo, Bunkyo-ku, Tokyo 113-0033, Japan}

\author{T.~Ari-izumi} 
\affiliation{Department of Physics, Konan University, Okamoto 8-9-1, Higashinada, Kobe 658-8501, Japan}

\author{F.~L.~Bello~Garrote}
\affiliation{Department of Physics, University of Oslo, N-0316 Oslo, Norway}
\affiliation{Department of Medical Physics, Oslo University Hospital, Oslo, Norway}

\author{L.~T.~Bell}
\affiliation{Department of Physics, University of Oslo, N-0316 Oslo, Norway}
\affiliation{Norwegian Nuclear Research Centre, Oslo, Norway}

\author{M.~M.~Bj{\o}r{\o}en}
\affiliation{Department of Physics, University of Oslo, N-0316 Oslo, Norway}

\author{F.~W.~Furmyr}
\affiliation{Department of Physics, University of Oslo, N-0316 Oslo, Norway}

\author{D.~Gjestvang}
\affiliation{Department of Physics, University of Oslo, N-0316 Oslo, Norway}
\affiliation{Norwegian Defence Research Establishment, NO-2027 Kjeller}

\author{A.~G{\"o}rgen}
\affiliation{Department of Physics, University of Oslo, N-0316 Oslo, Norway}
\affiliation{Norwegian Nuclear Research Centre, Oslo, Norway}

\author{V.~W.~Ingeberg}
\affiliation{Department of Physics, University of Oslo, N-0316 Oslo, Norway}
\affiliation{Norwegian Nuclear Research Centre, Oslo, Norway}

\author{K.~C.~W.~Li}
\affiliation{Department of Physics, University of Oslo, N-0316 Oslo, Norway}
\affiliation{Norwegian Nuclear Research Centre, Oslo, Norway}

\author{E.~Lima}
\affiliation{Department of Physics, University of Oslo, N-0316 Oslo, Norway}
\affiliation{Norwegian Nuclear Research Centre, Oslo, Norway}

\author{M.~Markova}
\affiliation{Department of Physics, University of Oslo, N-0316 Oslo, Norway}
\affiliation{Norwegian Nuclear Research Centre, Oslo, Norway}

\author{E.~F.~Matthews}
\affiliation{Department of Nuclear Engineering, University of California, Berkeley, California 94720, USA}

\author{A.~H.~Mj{\o}s}
\affiliation{Department of Physics, University of Oslo, N-0316 Oslo, Norway}
\affiliation{Norwegian Nuclear Research Centre, Oslo, Norway}

\author{S.~Miyamoto}
\affiliation{Laboratory of Advanced Science and Technology for Industry, University of Hyogo, 3-1-2 Kouto, Kamigori, Ako-gun, Hyogo 678-1205, Japan}

\author{V.~Modamio}
\affiliation{Department of Physics, University of Oslo, N-0316 Oslo, Norway}
\affiliation{Norwegian Nuclear Research Centre, Oslo, Norway}

\author{T.~Renstr\o m}
\affiliation{Department of Physics, University of Oslo, N-0316 Oslo, Norway}
\affiliation{Expert Analytics AS, N-0179 Oslo, Norway}

\author{E.~Sahin}
\affiliation{Department of Physics, University of Oslo, N-0316 Oslo, Norway}
\affiliation{Norwegian Nuclear Research Centre, Oslo, Norway}

\author{S.~Siem}
\affiliation{Department of Physics, University of Oslo, N-0316 Oslo, Norway}
\affiliation{Norwegian Nuclear Research Centre, Oslo, Norway}

\author{M.~Wiedeking}
\affiliation{Nuclear Science Division, Lawrence Berkeley National Laboratory, Berkeley, CA 94720, USA}
\affiliation{SSC Laboratory, iThemba LABS, P.O. Box 722, Somerset West 7129, South Africa}
\affiliation{School of Physics, University of the Witwatersrand, Johannesburg 2050, South Africa}

\date{\today}
 
\begin{abstract}
We present a study on the $\gamma$-decay properties of $^{64}$Zn using the Oslo method on $^{64}$Zn($p,p^\prime \gamma$) data combined with $^{64}$Zn$(\gamma,n)$ cross-section measurements at the NewSUBARU facility. 
With the Oslo method, we have measured the $\gamma$-ray strength function ($\gamma$SF) and the nuclear level density (NLD) below the neutron threshold. 
We observe that the NLD trend in the quasi-continuum region of $^{64}$Zn is best characterized by a constant-temperature-like model. 
Surprisingly, we find that $\gamma$-ray transitions from the quasi-continuum decaying directly to the $0^+$ ground state seem to be strongly hindered with a hindrance factor of $\kappa \approx 0.5$, 
which could be an indication of non-statistical effects in the ground-state decay due to, \textit{e.g.}, differences in nuclear shapes. 
For $\gamma$ energies above the neutron separation energy, the NewSUBARU ($\gamma, n$) data set probes a significant part of the giant dipole resonance. 
Furthermore, we find that the Oslo-method $\gamma$SF shows a rather smooth behavior, with a clear low-energy enhancement (LEE) for $E_{\gamma} <  4$~MeV.
\end{abstract}

\maketitle

\section{Introduction}
Atomic nuclei are extremely complex many-body systems, due to the nature of the short-ranged, strong nuclear force that keeps the neutrons and protons together in a very small volume. 
Nuclei are found to exhibit both collective behavior, such as rotation and vibration, and single-(quasi) particle behavior apparent from, \textit{e.g.}, pick-up and stripping reaction cross sections. 
In this respect, $^{64}$Zn is a very interesting case to study, as zinc isotopes have only two protons outside the $Z=28$ closed shell and are characterized by low-lying second and third excited $0^+$ states. 
Using a combination of $\gamma$-ray, conversion-electron, and internal-pair spectrometry on $^{64}$Zn, Passoja \textit{et al.}~\cite{passoja1984} were able to provide absolute $E0$ and $E2$ transition probabilities for many of the positive-parity states. 
They concluded that $^{64}$Zn (and the even-mass Zn isotopes in general) cannot be treated as typical vibrational
nuclei, and that the structure of the $0_2^+$ level in $^{64}$Zn  differs substantially from the structure of the ground-state band. 
Moreover, in a previous work by Druce \textit{et al.}~\cite{druce1982}, 
enhanced $E2$ transition rates of 20--40 W.u. were observed between ground-band members and between states built on the second $2^+$ state. 
This hints that the situation is rather complex with rotational-like bands built on bandheads of very different structures. 

Coulomb excitation experiments~\cite{koizumi2003} show that the $2^+$ state of the $^{66}$Zn ground-state band has a positive quadrupole moment, which indicates a quasi-rotational band of soft triaxial deformation. 
Recent work~\cite{rocchini2021} points out that the onset of triaxiality  is incompatible with simple collective models. The triaxiality of the zinc isotopes seems to be present all the way up to $^{72}$Zn. Here, safe Coulomb excitation in inverse
kinematics~\cite{hellgartner2023} shows that the $^{72}$Zn ground state is moderately deformed and with an average $\gamma$-degree of freedom close to maximum triaxiality ($\gamma \approx 30 ^{\circ}$). 

The well-known magnetic-dipole scissors mode~(see Ref.~\cite{Heyde2010} and references therein) has so far only been observed for prolate nuclear shapes in the quasi-continuum region, see \textit{e.g.}~Ref.~\cite{renstrom2018_Dy}. 
It is therefore an open question if such strong $M1$ decay branches could be present for triaxial shapes, or possibly more rigid oblate shapes. 
It is also interesting to investigate the impact of triaxiality in the $\gamma$-ray decay probability from the quasi-continuum energy regime \cite{Grosse2014}.

In the excitation-energy region with high level densities, it is convenient to work with groups of levels defined within certain energy bins. 
Thus, the {\em nuclear level density} (NLD) of available levels with spin $J$ and parity $\pi$ at a given excitation energy $E_x$ is given by
\begin{equation}
\rho(E_x,J,\pi)=\frac{\Delta N(E_x,J,\pi)}{\Delta E_x},
\end{equation} 
where $\Delta N$ is the number of levels within the excitation-energy bin $\Delta E_x$. 
By summing this expression over all spins and both parities, we obtain the total NLD $\rho(E_x)$. 
Furthermore, the $\gamma$SF describes the {\em average, reduced} $\gamma$-ray transition probabilities between groups of levels and is defined by~\cite{Bartholomew1973} 
\begin{equation}
f_{\rm XL}(E_{\gamma},E_i,J_i,\pi_i)=\frac{\langle \Gamma(E_i,J_i,\pi_i,E_{\gamma} ) \rangle}{E_{\gamma}^{2L+1}} \rho(E_i,J_i,\pi_i),
\end{equation}
where the level density $\rho$ is taken at the initial excitation energy $E_i$ for levels of spin $J_i$ and parity $\pi_i$.  
The $\gamma$SF is proportional to the average, partial $\gamma$-ray width $\langle \Gamma(E_i,J_i,\pi_i,E_{\gamma} ) \rangle$ of the group of initial levels, and the transition's $\gamma$ energy is given by $E_{\gamma}=E_i-E_f$ where $E_f$ is the final excitation-energy bin. 
The transitions of electromagnetic character $X$ and multipolarity $L$ obey the selection rules determined by the spins and parities of the initial and final levels.

There are several techniques to measure the NLD. 
At low excitation energies,  known discrete levels~\cite{NNDC} can be used to determine the level density. However, there are often missing levels already at a few MeV of excitation energy. 
A well-tested method is to extract the NLD from particle evaporation spectra~\cite{ soltesz2021}. Furthermore, the partial level density  at the neutron separation energy $S_n$ can be obtained from $s$-wave ($\ell=0$) neutron capture resonance spacings $D_0$~\cite{Mughabghab2018}. 
At higher excitation energies, Ericson fluctuations~\cite{ericson1960} can provide information on the NLD. 

Turning to the $\gamma$SF,  nuclear resonance fluorescence (NRF) experiments~\cite{savran2013,zilges2022} have been the workhorse for retrieving $\gamma$-transition strengths below the neutron separation energy. 
Direct and average resonance capture data~\cite{IAEA2022} have also frequently been used to extract the $\gamma$SF. 
Furthermore, the two-step cascade~\cite{Krticka2004} and the multi-step cascade (MSC) method~\cite{valenta2017} can be used to reveal structures in the $E_{\gamma} = 2-4$~MeV region. 
A compilation of these methods was provided in Ref.~\cite{Goriely2019}, and $\gamma$SF data from many different measurements have been published in a database~\cite{IAEA2022} provided by the International Atomic Energy Agency (IAEA).

The Oslo method~\cite{Gut87,Gut96,Schiller00} is 
an experimental technique that allows for a simultaneous extraction of the NLD and $\gamma$SF in one and the same experiment. 
Recently, the shape method~\cite{Wiedeking2021} has been applied to further exploit the same measured data set with an independent analysis. 
In this work, we will apply both methods, and also present data on the $\gamma$SF above the neutron separation energy from $(\gamma,n)$ cross-section data measured at the NewSUBARU facility~\cite{NewSUBARU2009}.

The manuscript is outlined as follows. The experimental set-up and techniques applied at the NewSUBARU and Oslo laboratories are described in Sec.~II. 
In Sec.~III, the NLD and hindrance of direct transitions from the quasi-continuum to the ground level are discussed, and the experimental NLD is compared to theory predictions from large-scale shell-model calculations. 
The  $\gamma$SF results from the Oslo data are provided in Sec.~IV. 
Section~V presents the experimental ${\gamma}$SF of $^{64}$Zn from all the considered data sets, and a comparison with semi-empirical models. 
A summary and outlook are given in Sec.~VI.

\section{Experiments}
The present work is based on two different experimental techniques. 
The Oslo-type experiment, performed at the Oslo Cyclotron Laboratory (OCL), University of Oslo, Norway, is based on charged-particle reactions to extract the NLD and $\gamma$SF for energies below $S_n$. 
The Oslo method has been extensively tested and discussed in Ref.~\cite{Lars11}, and the data-analysis software is available on the OCL GitHub~\cite{om2022}.

At higher energies, we have utilized the $(\gamma,n)$ reaction from quasimonochromatic laser Compton-scattered $\gamma$ rays produced at the NewSUBARU facility operated by the University of Hyogo, Japan. 
The cross-section from these measurements is translated to $\gamma$SF following Axel's prescription~\cite{Axel1968} for comparison with the Oslo data. 
Using the two techniques, our aim is to obtain an experimental $\gamma$SF for a wide range of $\gamma$-ray energies, $E_\gamma \approx$ 2.0--20~MeV. 

\subsection{The NewSUBARU experiment}
The photoneutron cross-section measurement on $^{64}$Zn took place at the NewSUBARU synchrotron radiation facility in Japan. 
Quasi-monochromatic $\gamma$-ray beams were obtained through laser Compton scattering (LCS) of 1064 nm photons in head-on collisions with relativistic electrons.
The $\gamma$ beams were collimated with Pb collimators.  
Throughout the experiment, the laser was periodically on for 80 ms and off for 20 ms to measure background neutrons and $\gamma$ rays. 
In this experiment, the beams produced had an energy resolution ranging from 0.6 MeV to 0.9 MeV FWHM.

The electrons were injected from a linear accelerator into the NewSUBARU storage ring with an initial energy of 1085 MeV, then subsequently decelerated to  
provide LCS $\gamma$-ray beams of energies in the range of 11.94--20.92 MeV. 
The maximum $\gamma$-ray energy of the beams was increased in steps of 0.25 MeV. 
The electron beam energy has been calibrated with an accuracy on the order of 10$^{-5}$ \cite{calofns}.  

The energy profiles of the produced $\gamma$-ray beams were measured with a $\diameter$3.5 in.$\times$4 in. LaBr$_3$(Ce)  detector. The measured LaBr$_3$(Ce) spectra were reproduced by a \textsf{Geant4} code~\cite{filipescu2022, filipescu2023, geant1, geant2, geant3} that incorporated the kinematics of the LCS process, including the beam emittance and the interactions between the LCS beam and the LaBr$_3$(Ce) detector. 

The target sample from Trace Sciences was 1.00 g of solid metallic $^{64}$Zn, enriched to 99.4(1)\% and placed inside an aluminum cylinder. 
Due to the collimation of the beam, the $\gamma$ beam did not hit the aluminum housing. 
Also, the container did not have an aluminum cap, allowing for measurements of the photoneutron cross section well above the neutron separation energy of $^{27}$Al.

To measure the emitted neutrons, a high-efficiency $4\pi$ detector was used, consisting of 20 $^3\rm{He}$ proportional counters, arranged in three concentric rings and embedded in a 36 $\times$ 36 $\times$ 50 cm$^3$ polyethylene neutron moderator~\cite{itoh_2011}. 
The ring ratio technique, originally developed by Berman and Fultz~\cite{Berman_ring_ratio}, was used to determine the average energy of the neutrons from the ($\gamma,n$) reactions. 
The efficiency of the neutron detector varies with the average neutron energy. 
The efficiency was measured with a calibrated $^{252}$Cf source with the emission rate of 2.27 $\times$ 10$^4$ s$^{-1}$ with 2.2\% uncertainty, and the energy dependence was determined by Monte Carlo simulations \cite{Nyhu15}. 
The efficiency of the neutron detector was simulated using isotropically distributed, mono-energetic neutrons. 
Once the neutron detection efficiency for a given beam energy has been determined, we were able to deduce the number of ($\gamma, n$) reactions that took place during each run. 

The LCS $\gamma$-ray flux was monitored by a  $\diameter$8 in.$\times$12 in. NaI(Tl) detector during neutron measurement runs with 100$\%$ detection efficiency for the beam energies used in this experiment. The number of incoming $\gamma$ rays per measurement was determined using the pile-up and Poisson-fitting technique described in Refs.~\cite{KONDO2011462,UTSUNOMIYA2018103}.

\begin{figure}[t]
    \includegraphics[clip,width=1.0\columnwidth]{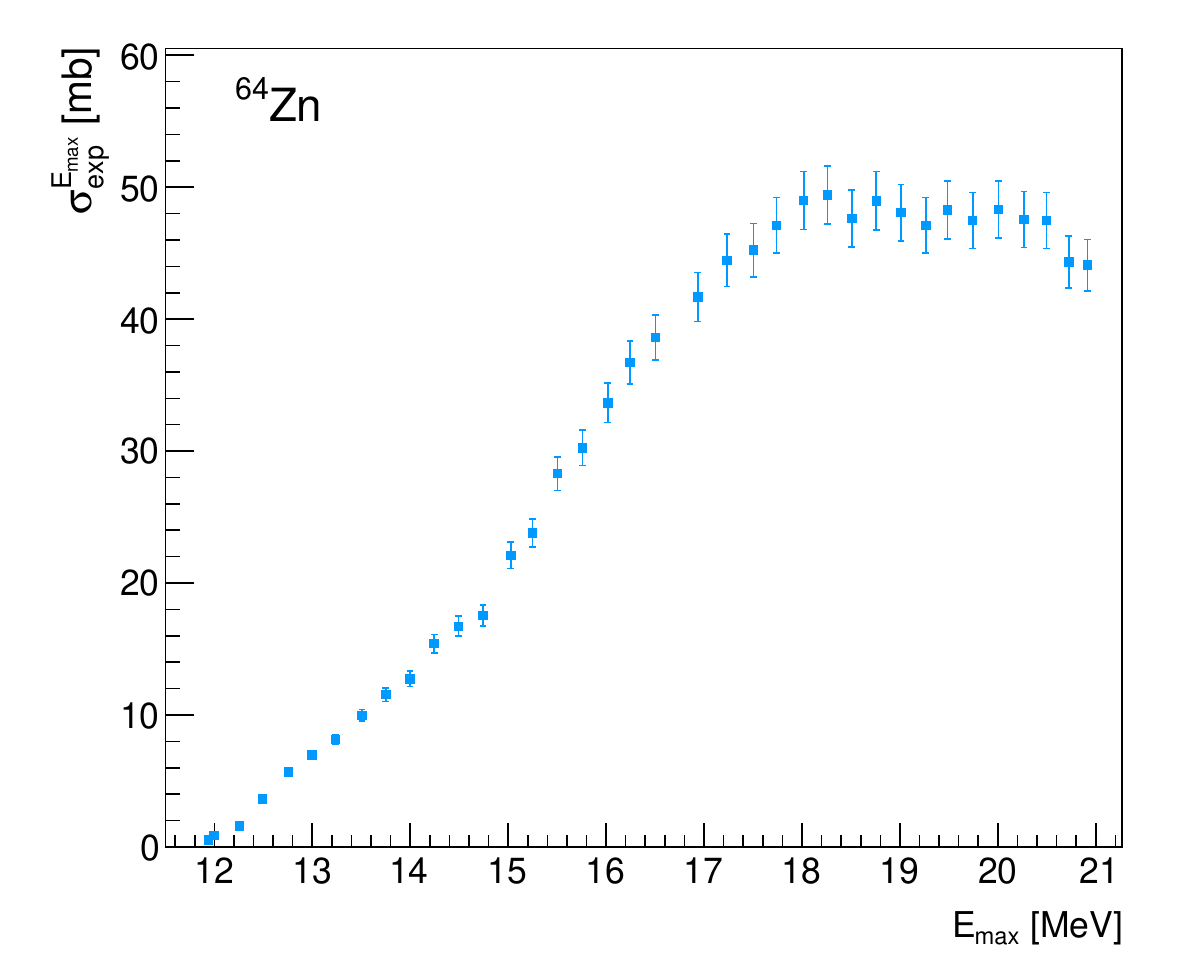}
    \caption{(Color online) The convoluted cross section $\sigma^{E_{\rm max}}_{\rm exp}$ vs. the maximum $\gamma$-beam energy $E_{\rm max}$ (see text). }
    \label{fig:mono}
\end{figure}

The measured photo-neutron cross section for an incoming beam with maximum $\gamma$ energy $E_{\rm max}$ is given by the convoluted cross section,
\begin{equation}
\sigma^{E_{\rm max}}_{\rm exp}=\int_{S_n}^{E_{\rm max}}D^{E_{\rm max}}(E_{\gamma})\sigma(E_{\gamma})dE_{\gamma}=\frac{N_n}{N_tN_{\gamma}\xi\epsilon_n g}.
\label{eq:cross1}
\end{equation}
Here, $D^{E_{\rm max}}$ is the normalized energy distribution ($\int_{S_n}^{E_{\rm max}} D^{E_{\rm max}}dE_{\gamma}= 1$) of
the $\gamma$-ray beam obtained from \textsf{Geant4} simulations. 
Furthermore, $\sigma(E_{\gamma})$ is the true photo-neutron cross section as a function of $\gamma$ energy. 
The quantity $N_n$ represents the number of neutrons detected, $N_t$ gives the number of target nuclei per unit area, $N_{\gamma}$ is the number of $\gamma$ rays incident on the target sample, $\epsilon_n$ represents the neutron detection efficiency, and  $\xi=(1-e^{-\mu t})/(\mu t)$ gives the correction factor for self-attenuation in the target sample. 
Finally, the factor $g$ represents the fraction of the $\gamma$ flux above $S_n$. 
The convoluted cross section $\sigma^{E_{\rm max}}_{\rm exp}$ for $^{64}$Zn is shown in Fig.~\ref{fig:mono}.
As seen from Eq.~(\ref{eq:cross1}), to obtain an estimate for the true cross-section $\sigma(E_{\gamma})$, we need to deconvolute, or \textit{unfold}, the measured cross section, taking into account the energy profile of the $\gamma$ beams.
By approximating the integral in Eq.~(\ref{eq:cross1}) with a sum for each $\gamma$-beam profile, the unfolding problem can be expressed as a set of linear equations:
\begin{equation}
\sigma_{\rm f }=\bf{D}\sigma,
\end{equation}
where $\sigma_{\rm f}$ is the cross section folded with the beam profile {\bf D}.  
The indices $i$ and $j$ of the matrix element $D_{ij}$ correspond to $E_{\rm max}$ and $E_{\gamma}$, respectively.
The set of equations is given by
\begin{equation}
\begin{pmatrix}\sigma_{\rm{1f}}\\\sigma_{\rm{2f}}\\ \vdots \\ \sigma_{N{\rm f}} \end{pmatrix}\\\mbox{}=
\begin{pmatrix}D_{ 11} & D_{ 12} & \cdots &\cdots &D_{ 1M} \\ D_{ 21} & D_{ 22} &
\cdots & \cdots & D_{ 2M} \\ \vdots & \vdots & \vdots & \vdots & \vdots \\ D_{ N1} & D_{ N2}& \cdots & \cdots &D_{ NM}\end{pmatrix}
\begin{pmatrix}\sigma_{1}\\\sigma_{2}\\ \vdots \\ \vdots \\\sigma_{M} \end{pmatrix}.
\label{eq:matrise_unfolding}
\end{equation}
Each row of $\bf{D}$ corresponds to a \textsf{Geant4} simulated $\gamma$
beam profile belonging to a specific measurement characterized by $E_{\rm max}$.  
As the system of linear equations in Eq.~(\ref{eq:matrise_unfolding}) is under-determined, the true $\sigma$ vector cannot be estimated by matrix inversion. 
To find $\sigma$, we utilize the folding iteration method described in Refs.~\cite{renstrom2018_Dy,larsen2023}.

\begin{figure}[t]
    \includegraphics[clip,width=1.0\columnwidth]{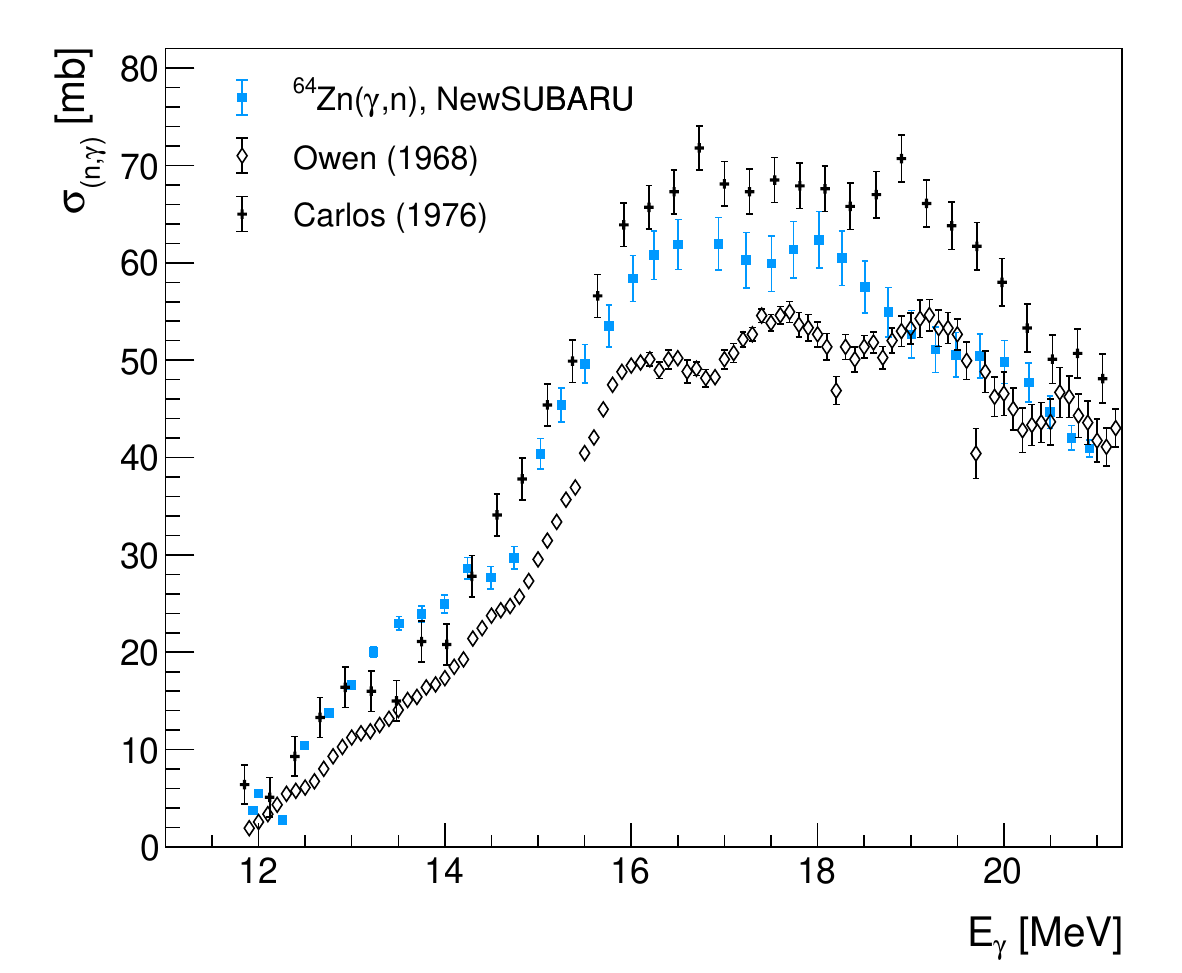}
    \caption{(Color online)  Unfolded data from this work (azure filled circles) compared to the results from Owen \textit{et al.}~\cite{owen1968} (open diamonds) and Carlos \textit{et al.}~\cite{carlos1976} (black crosses). }
    \label{fig:cross_64Zn}
\end{figure}
The unfolded data are shown in Fig.~\ref{fig:cross_64Zn} and compared to previous $(\gamma,n)$ measurements by Owen \textit{et al.}~\cite{owen1968} and Carlos \textit{et al.}~\cite{carlos1976}.
We observe that our data follow rather closely the results of Carlos \textit{et al.} up to $E_\gamma \approx 15.5$ MeV, while the data points from Owen \textit{et al.} are significantly lower in absolute value. 
We also notice that we do not observe a third peak in the cross section at $E_\gamma \approx 19.2$ MeV as seen in the previous measurements. 

\subsection{The Oslo experiment}
\begin{figure*}[t]
\includegraphics[clip,width=2.\columnwidth]{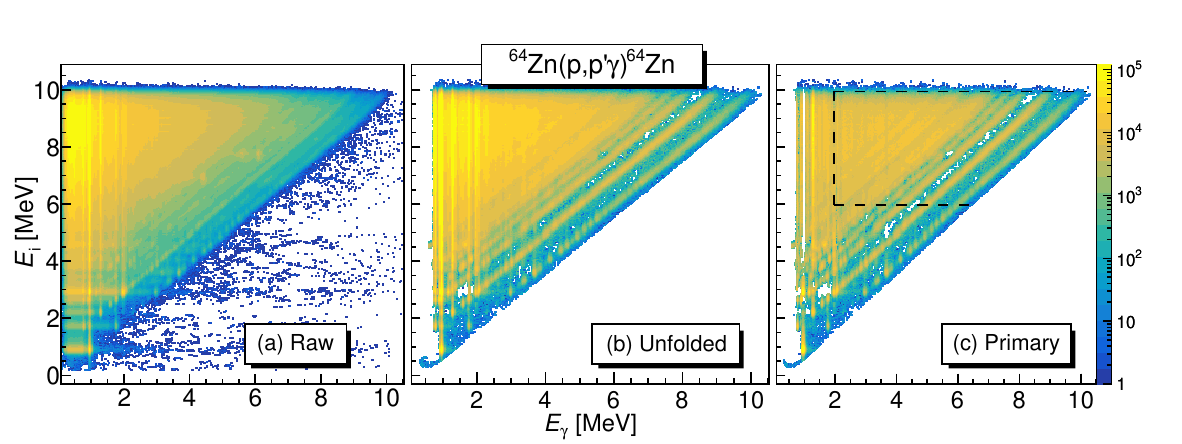}
\caption{(Color online) Proton-$\gamma$ coincidence matrices of the $^{64}$Zn($p,p'\gamma $)$^{64}$Zn reaction, where the deposited proton energy in SiRi is converted to initial excitation energy $E_i$. 
The raw (a), unfolded  (b), and primary matrices (c)  show $\gamma$ spectra as function of initial excitation energies $E_i$. 
The primary matrix $P(E_{\gamma},E_i)$ (c) represents the starting point for the Oslo method. The region within the dashed lines is used for the subsequent analysis. The $E_{\gamma}$ and $E_{i}$ energy bins each have widths of 36 keV. }
\label{fig:matrices}
\end{figure*}

The $^{64}$Zn isotope was studied with the $(p,p'\gamma)$ reaction with a proton beam energy of 16~MeV at the Oslo Cyclotron Laboratory. 
The target was a self-supporting metallic foil with a thickness of $\approx 2$~mg/cm$^2$ and enriched to $\approx 97$\% in $^{64}$Zn. 

The SiRi particle-telescope system~\cite{siri} was used to measure the outgoing charged particles. 
The particle telescopes were placed at a distance of~5 cm from the target. The segmented front detectors covered eight angular bins from $126^{\circ}$ to $140^{\circ}$ with respect to the beam direction. 
The front ($\Delta E$) and back ($E$) detectors had thicknesses of 130 and 1550 $\mu$m, respectively.
The obtained energy resolution for this experiment was $\approx$ 150 keV (full width at half maximum, FWHM) for the total deposited particle energy ($E + \Delta E$). 
The coincident $\gamma$ rays were recorded with the OSCAR array~\cite{Zeiser2021,Goergen2021}, which consists of 30 LaBr$_3$(Ce) scintillators. 
The cylindrical crystals ($\diameter$3.5 in. $\times$ 8 in.) were mounted spherically around the target at a distance of 16.3~cm.

The first step of the Oslo method is to sort the coincidence events into $\gamma$ spectra for each initial excitation energy $E_i$, which is determined from the total deposited energy in SiRi and the reaction kinematics.  
Figure \ref{fig:matrices}(a) shows this unprocessed (``raw''), background-subtracted $(E_{\gamma}, E_i)$ matrix. 
Each $\gamma$ spectrum is then unfolded using the known detector response functions for OSCAR~\cite{om2022,Zeiser2021}. 
Some small contaminant contributions from  $\gamma$ transitions in $^{12}$C and $^{16}$O were removed with a 2D Gaussian peak-fit procedure after the unfolding. 
The matrix obtained from the unfolding procedure~\cite{Gut96} is displayed in Fig.~\ref{fig:matrices}(b), which presents all $\gamma$ rays included in the cascades from each $E_i$.

Next, we need to determine the primary $\gamma$-ray matrix based on the unfolded spectra $u_{E_i}(E_{\gamma})$. 
The primary spectrum at initial excitation energy $E_i$ can be obtained by subtracting a sum of weighted $u_{E_i'}(E_{\gamma})$ spectra from lower excitation energies $E_i'$:
\begin{equation}
p_{E_i}(E_{\gamma})=u_{E_i}(E_{\gamma}) - \sum_{E_i' < E_i}w_{E_i}(E_i')u_{E_i'}(E_{\gamma}),
\end{equation}
where the weighting coefficients $w_{E_i}(E_i')$ and primary $\gamma$ spectrum $p_{E_i}(E_{\gamma})$ are determined by a fast converging iterative procedure~\cite{Gut87}. 
The obtained primary $P(E_{\gamma},E_i$) matrix for  $^{64}$Zn is shown in Fig.~\ref{fig:matrices}(c). 
For more details, see Appendix~\ref{appC}.

To extract the NLD and $\gamma$SF, we first normalize the primary spectra to unity for each $E_i$ bin by $\sum_{E_{\gamma}}P(E_{\gamma}, E_i)=1$ and then factorize $P$ by~\cite{Schiller00}
\begin{equation}
P(E_{\gamma}, E_i) \propto   \rho(E_i-E_{\gamma}){\cal{T}}(E_{\gamma}) ,\
\label{eqn:rhoT}
\end{equation}
where we assume that the decay probability is proportional to the NLD at the final excitation energy $\rho(E_i-E_{\gamma})$ according to Fermi's golden rule~\cite{dirac,fermi}. 
The decay is also proportional to the $\gamma$-ray transmission coefficient ${\cal{T}}$, which is assumed to be independent of excitation energy, spin and parity according to the Brink-Axel hypothesis~\cite{brink,Axel1962}.

Provided that the relation given in Eq.~(\ref{eqn:rhoT}) holds, we can extract the one-dimensional vectors $\rho$ and ${\cal{T}}$ from the two-dimensional $P$ matrix using an iteration procedure based on a least-$\chi ^2$ fit on the experimental primary $\gamma$ spectra, see Ref.~\cite{Schiller00}. 
We restrict the fit to the part of the $P$ matrix where the decay can be considered statistical, i.e., where decay of compound-nucleus states dominate. 
In practice, this means that the excitation energy needs to be sufficiently high, ensuring that the density of the initial levels is so high that they are strongly mixed and decay in a statistical manner.
Further, a proper extraction of the primary $\gamma$ spectra is crucial. 
From the convergence tests and tests of the correctness of the primary spectra obtained from the iterative procedure~\cite{Gut87}, we find that the excitation energy region of 6.0~MeV~$< E_i < $~10~MeV with $E_{\gamma}> 2.0$~MeV works well for $^{64}$Zn (see also Appendix~\ref{appC}).  
\begin{figure}[t]
\begin{center}
\includegraphics[clip,width=1.\columnwidth]{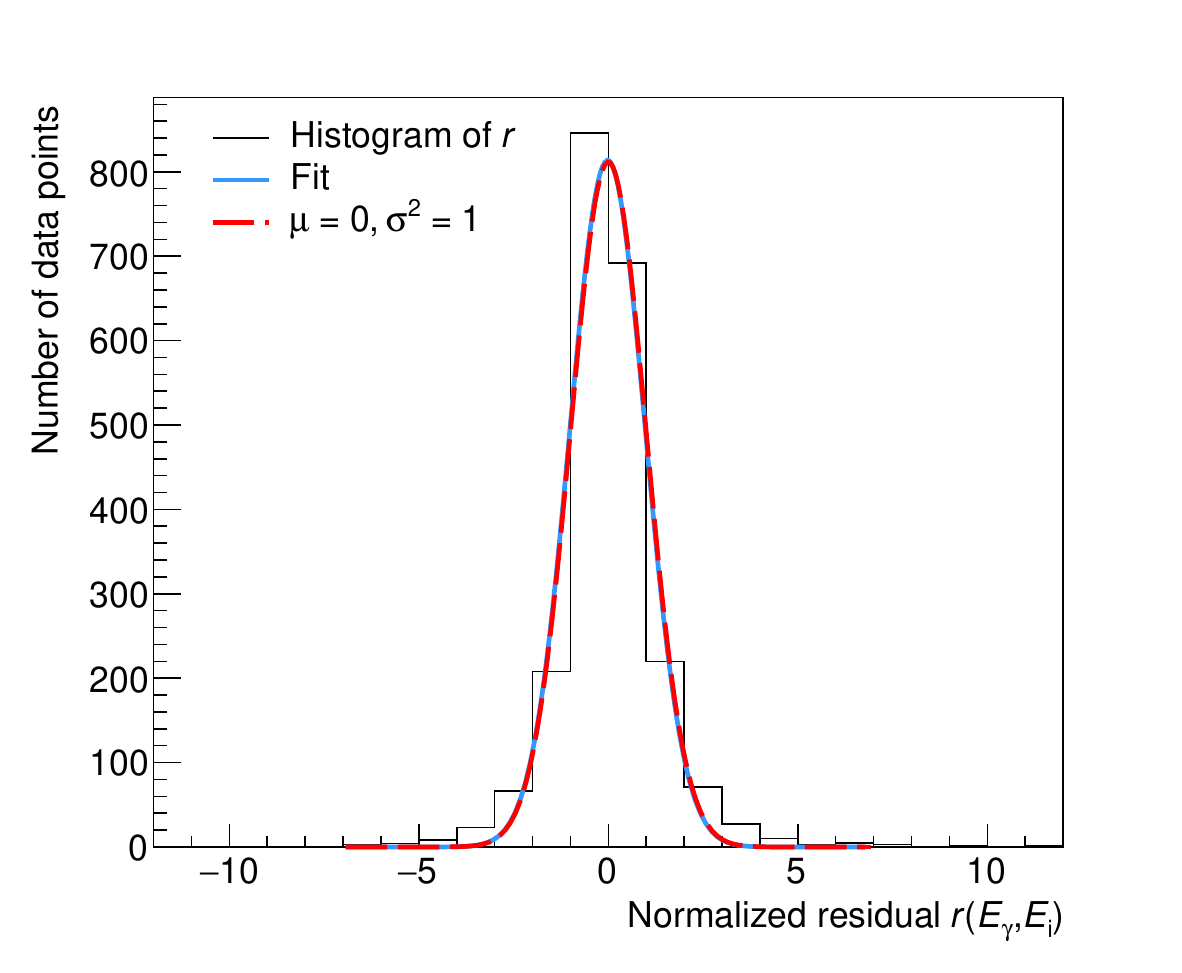}
\caption{(Color online) Distribution of residuals after multiplying the originally deduced uncertainties with a common factor of 5.0, see text and Eq.~(\ref{eq:residuals}). The azure, solid line is a fit to the histogram, while the red, dashed line shows for comparison a normal distribution with $\mu = 0$ and $\sigma^2 = 1$. }
\label{fig:residuals}
\end{center}
\end{figure}
\begin{figure*}[t]
\begin{center}
\includegraphics[clip,width=1.9\columnwidth]{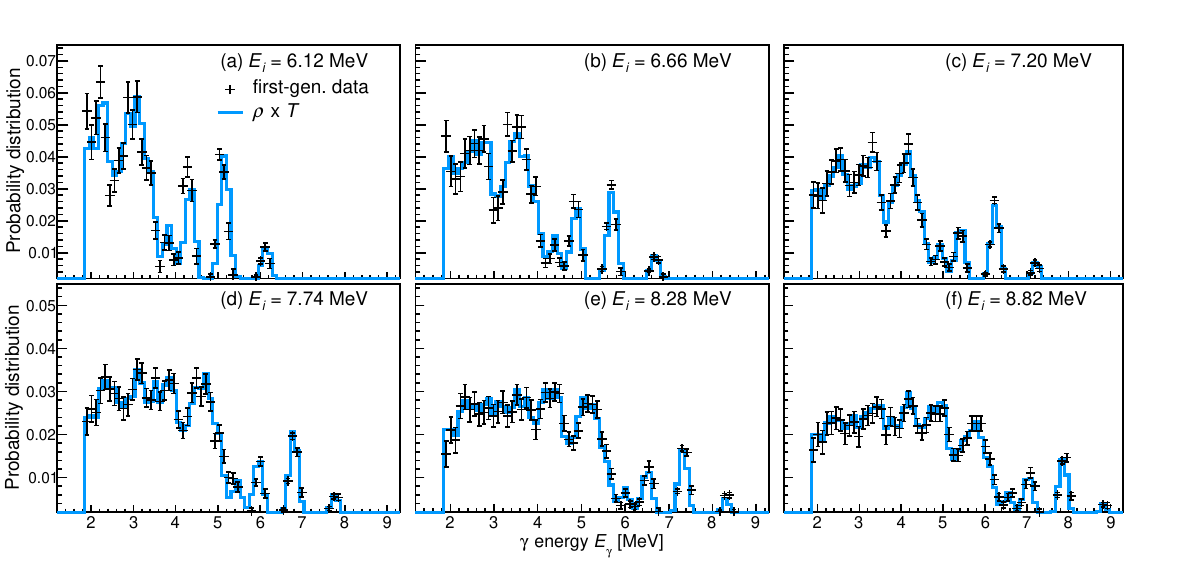}
\caption{(Color online) Probability density distributions of primary $\gamma$ spectra (crosses) as function of initial excitation energies $E_i$ for $^{64}$Zn. The spectra are compared to the product $\rho(E_i-E_{\gamma}) {\mathcal{T}}(E_{\gamma})$ (blue histogram). The estimated uncertainties are one-standard deviation (68.3\% confidence level) after applying the common scaling factor of 5.0 on the original estimates for the experimental uncertainties. The $E_{\gamma}$ and $E_{i}$ energy bins have widths of 108 keV.}
\label{fig:doesitwork}
\end{center}
\end{figure*}

Estimating uncertainties for the $\rho(E_i-E_\gamma)$ and ${\cal{T}}(E_\gamma)$ vectors is not a trivial task, as there are several sources of error that are difficult to quantify.
In particular, errors in the response functions of OSCAR and potential problems in the iterative method to obtain primary spectra are difficult to deduce. 
In addition, Porter-Thomas fluctuations~\cite{Porter1956} are expected to be significant as $^{64}$Zn is a relatively light nucleus with a low level density.
As a first approximation, we have estimated the uncertainties according to the procedure described in Ref.~\cite{Schiller00}. 
However, with these experimental uncertainties, we obtained a large value of the reduced $\chi^2$, $\chi^2_{\mathrm{red}} \approx 47$, from the comparison of the ``theoretical'' primary spectra from the product $\rho(E_i-E_{\gamma}){\cal{T}}(E_{\gamma})$ with the experimental primary $\gamma$ spectra.
Further, when plotting the normalized residuals of the experimental primary $\gamma$ spectra versus the frequency of the residuals, we find a distribution that is centered around zero, but with a standard deviation that is much bigger than unity. 
As discussed by Pruitt \textit{et al.}~\cite{Pruitt2023}, this is a signature of the experimental uncertainties being underestimated. 
To improve this situation, we have applied the inflation factor method (also called the scale factor method) as discussed in the work of Adelberger \textit{et al.}~\cite{Adelberger2011}.
In this method, all the experimental errors are scaled by a common factor. 
In our case, we find that a common scaling factor of 5.0 applied on the originally estimated uncertainties gives a distribution of the normalized residuals that is in fair agreement with a normal distribution with mean $\mu = 0$ and a standard deviation $\sigma = 1$, see Fig.~\ref{fig:residuals}.
Here, we calculate the normalized residuals $r(E_\gamma,E_i)$ according to 
\begin{equation}  
r(E_\gamma,E_i) = \frac{ P(E_\gamma,E_i)_{\mathrm{exp}}-P(E_\gamma,E_i)_{\mathrm{th}}}{\Delta P(E_\gamma,E_i)_{\mathrm{exp}}},
\label{eq:residuals}
\end{equation} 
for all $E_\gamma$ and $E_i$ within the region marked in Fig.~\ref{fig:matrices}(c), and with $\Delta P(E_\gamma,E_i)_{\mathrm{exp}}$ being the re-evaluated uncertainties.

With the re-evaluated uncertainties, we compare the experimental primary $\gamma$ spectra and the results using the product of the two functions $\rho$ and ${\cal T}$. 
Figure~\ref{fig:doesitwork} shows six out of the 37 available $\gamma$ spectra that have been used to determine $\rho$ and ${\cal{T}}$ in the fitting procedure. 
We observe that the overall agreement is good, and that only a few data points are off by more than two standard deviations (the uncertainties shown in the figure are the estimated point-by-point one-standard deviation uncertainties). 

The local variations in the data points of $\rho$ and $\cal{T}$ are uniquely determined through the fit, but the scale and slope of these functions are still undetermined. It has been shown that transformations of the type~\cite{Schiller00}
\begin{eqnarray}
\tilde{\rho}(E_i-E_\gamma)&=&A\exp[\alpha(E_i-E_\gamma)]\,\rho(E_i-E_\gamma),
\label{eq:array1}\\
\tilde{{\mathcal{T}}}(E_\gamma)&=&B\exp(\alpha E_\gamma){\mathcal{T}} (E_\gamma),
\label{eq:array2}
\end{eqnarray}
give identical fits to the primary $\gamma$ spectra. The scaling parameters $A$ and $B$, and the slope parameter $\alpha$ are usually determined from other experimental data or systematics.
As mentioned in the Introduction, the shape method can also in principle be applied on the Oslo-type data set to determine the slope parameter $\alpha$. 
However, in this case it turns out that for $^{64}$Zn the shape method cannot be applied without some significant modifications, as will be discussed in Sec.~\ref{sec:shapemethod}.

\subsection{Normalization of the NLD and $\gamma$SF}
\label{sec:nldnorm}
\begin{figure}[t]
    \includegraphics[clip,width=1.0\columnwidth]{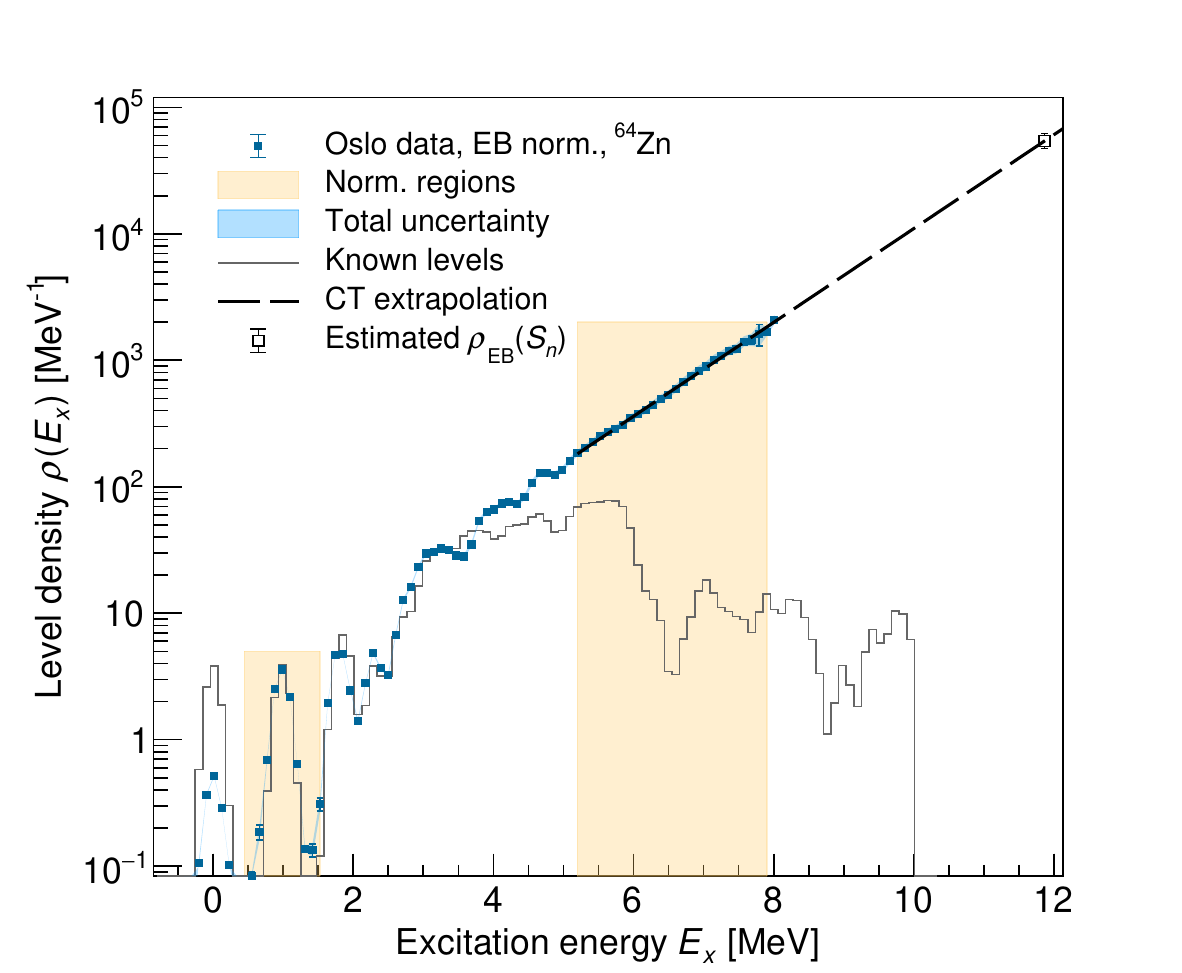}
    \caption{(Color online) Level density (filled data points) of $^{64}$Zn normalized to $\rho_{\mathrm{EB}}(S_n)$. 
    The data points are extrapolated by the dashed line to $\rho_{\mathrm{EB}}(S_n)$  using the constant-temperature model. 
    The histogram shows the level density of known levels~\cite{NNDC}, which has been smoothed with a Gaussian distribution with a FWHM of 230~keV. 
    The shaded boxes indicate the regions used for fitting to discrete levels  and the CT model.
    The energy bins have a width of 108 keV.}
    \label{fig:counting_EB}
\end{figure}

\begin{table*}[th]
\centering
\caption{Parameters used for normalizing the experimental NLD and $\gamma$SF for $^{64}$Zn. All uncertainties are given as $1\sigma$ confidence intervals. Here, $E_d$ is the discrete excitation energy for which the discrete spin-cutoff parameter $\sigma_d$ is estimated, $\sigma_J(S_n)$ is the spin-cutoff parameter at the neutron separation energy $S_n$, $a$ and $E_1$ are the level-density parameter and backshift, respectively, and $\rho(S_n)$ is the level density at $S_n$. Further, $D_0$ is the $s$-wave average level spacing, $T_{\rm CT}$ is the temperature parameter in the CT model, and $\langle\Gamma_{\gamma0}\rangle$ is the $s$-wave average, total radiative width. The calculated parameters for the GC~\cite{Gilbert1965} and EB~\cite{egidy2005} are obtained with the \textsf{Robin} code in the Oslo-method software package~\cite{om2022}.}
\begin{tabular}{ccccccccccc}
\hline
\hline
Model & $E_d$&$\sigma_d$ &$S_n$  & $\sigma_J(S_n)$&    $a$      &$E_1$& $\rho(S_n)$$^a$ & $D_0$$^b$ &$T_{\rm CT}$&$\langle\Gamma_{\gamma0}\rangle$$^c$\\
 & [MeV]&           &[MeV]  &              &[MeV$^{-1}]$ &[MeV]& [$10^{4}$ MeV$^{-1}]$ & [eV] & [MeV]   &           [meV] \\
\hline
GC & 3.0  &  2.4(2)   & 11.862&       3.60   &  7.827       &1.254 &   4.43$^{+0.67}_{-0.58}$& 175$^{+27}_{-23}$ & 1.20(2)  &    $708^{+298}_{-210}$ \\
EB & 3.0  &  2.4(2)   & 11.862&       4.38   &  7.380       &1.019 &   5.41$^{+0.78}_{-0.68}$& 200$^{+29}_{-25}$ & 1.17(2)  &    $708^{+298}_{-210}$ \\
\hline
\hline
\end{tabular}
\label{tab:gsf_parameters}

$^a$Estimated from semi-empirical systematics (see Appendix~\ref{appA}). \\
$^b$Estimated from $\rho(S_n)$ for the given spin-cutoff model.\\$^c$Estimated from $\Gamma_{\gamma0}$ data (see Appendix~\ref{appA}).
\end{table*}
The NLD extracted using Eq.~(\ref{eqn:rhoT}) must be normalized with respect to the scaling parameter $A$ and slope $\alpha$. 
For this purpose, we need two anchor points. 
At low excitation energies, we normalize the NLD to the known, discrete levels given in the database of the National Nuclear Data Center (NNDC)~\cite{NNDC}, smoothed with the experimental excitation-energy resolution ($\approx 230$ keV FWHM due to the combined resolution of the particle telescope and OSCAR). 
Specifically, we fit our data points to the first $2^+$ level at an excitation energy of $E=992$~keV as shown in Fig.~\ref{fig:counting_EB}. 
Also, for the excitation-energy region $E_x \approx$ 3.2--4.2 MeV, we note that some of the listed levels in the database of NNDC are indicated as uncertain and/or have no observed $\gamma$ ray. 
For the further comparison of our data with the discrete levels, we have excluded these dubious levels\footnote{Levels at $E_x =$ 3240, 3285, 3414, 3465, 3500, 3538.7, 3545.9, 3630, 3680, 3759, 3780, 3880, 4110, 4153.1 keV}.
At high excitation energies, we use the measured average neutron $s$-wave resonance spacing $D_0$~\cite{Mughabghab2018} at the neutron separation energy $S_n$ if available, to estimate the total level density at $S_n$, $\rho(S_n)$. 
However, as $^{63}$Zn is unstable, no $D_0$ value has been measured for $^{64}$Zn. 
We have therefore estimated the unknown $\rho(S_n)$ for $^{64}$Zn using the available $s$-wave and $p$-wave ($D_1$) neutron-resonance data on the other Zn isotopes, taken from Ref.~\cite{Mughabghab2018}.
For further details, we refer the reader to Appendix~\ref{appA}.

In principle, both the spin-cutoff parameter approaches of Gilbert and Cameron (GC)~\cite{Gilbert1965} and von Egidy and Bucurescu (EB)~\cite{egidy2005} are consistent with the experimental spin-distribution information that exists in this mass region~\cite{Grimes1974}. 
However, the EB spin-cutoff parameter leads to a spin distribution much closer to the one found from large-scale shell model calculations, as discussed in Sec.~\ref{sec:KSHELL}.
In addition, we find that the two approaches give overall rather similar results for the normalization.
Thus we proceed with the EB model in the following, and include the GC results in Appendix~\ref{appA} for completeness.

To exploit the second anchor point, we also need to extrapolate the experimental data up to $E_x=S_n$. 
We observe that the experimental NLD shows an approximately log-linear dependence in the region $E_x\approx 5.0 -8.0$~MeV, see Fig.~\ref{fig:counting_EB}. 
Hence, we  make use of the constant-temperature (CT) model of Ericson~\cite{Ericson59}:
\begin{equation}
\rho_{\text{CT}}(E_x)=(1/T_{\rm CT}) \exp{[(E_x-E_0)/T_{\text{CT}}],}
\label{eq:ct}
\end{equation}
 with two free parameters, the excitation-energy shift $E_0$ and the ``temperature'' $T$,
and extrapolate up to the anchor point $\rho(S_n)$. The parameters and values used in the normalization of $^{64}$Zn are listed in Table~\ref{tab:gsf_parameters}. 

From the normalized NLD, we have also determined the slope $\alpha$ of the $\gamma$-transmission coefficient as can be seen from Eqs.~(\ref{eq:array1}),~(\ref{eq:array2}). 
The absolute value given by the scaling parameter $B$ remains to be found. 
Again, as $^{63}$Zn is unstable, there is no experimental information on the average, total radiative width $\left< \Gamma_{\gamma0}\right>$ from $s$-wave neutron resonances, which is usually applied to normalize the $\gamma$-transmission coefficient \cite{Voinov2001}. 
Therefore, we make an estimate for the systematics of $\left< \Gamma_{\gamma0}\right>$ using the available $s$-wave resonance data for all the other Zn isotopes from Mughabghab's evaluation of neutron-resonance data~\cite{Mughabghab2018} (see Appendix~\ref{appA}).

\begin{figure}[t]
    \includegraphics[clip,width=1.0\columnwidth]{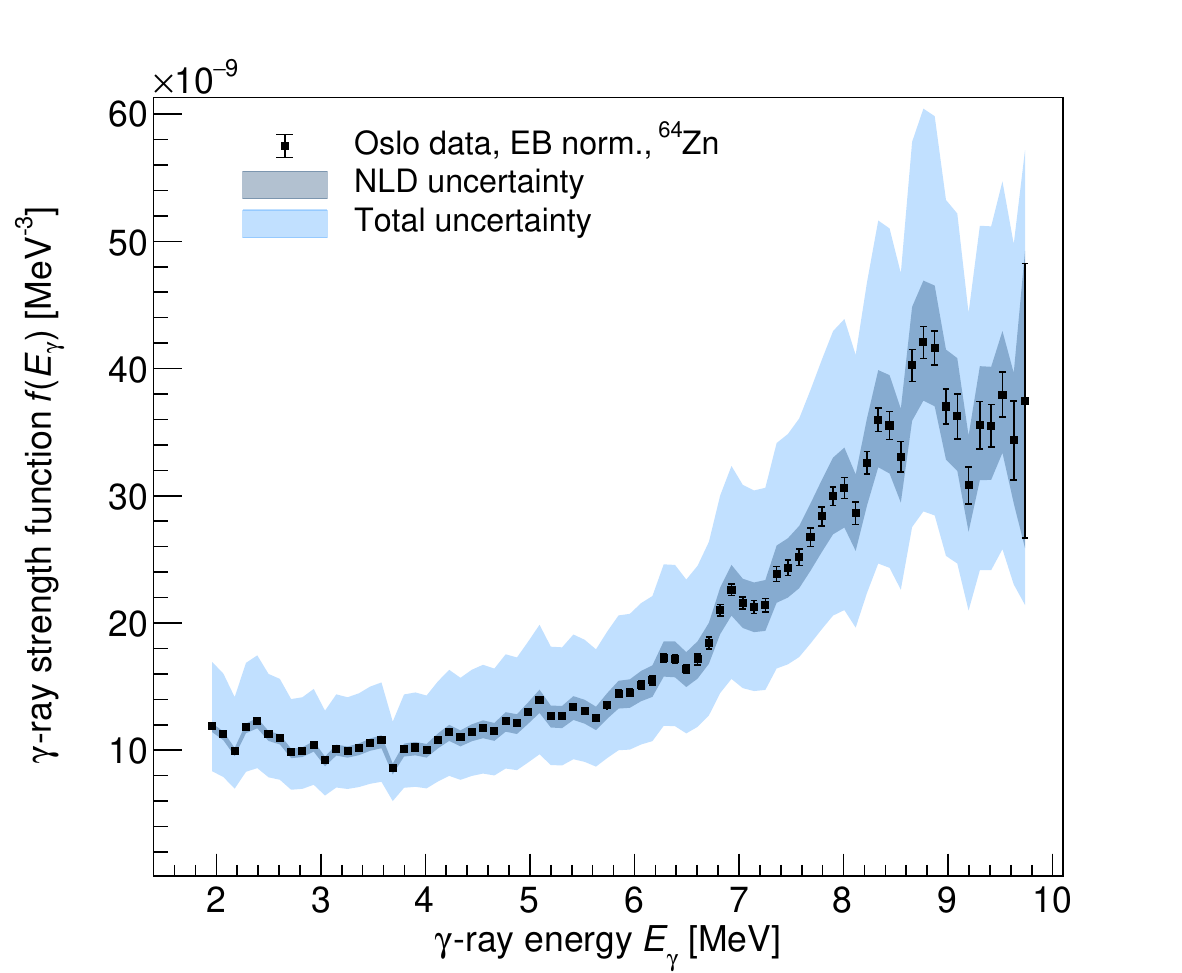}
    \caption{(Color online) Normalized $\gamma$SF of $^{64}$Zn for the EB normalization (see text). The energy bins have a width of 108 keV.}
    \label{fig:gsfEB}
\end{figure}

With the transmission coefficient ${\cal{T}}$ at hand, we obtain the dipole $\gamma$SF by~\cite{kopecky1990}
\begin{equation}
f (E_{\gamma}) =\frac{1}{2 \pi}\frac{{\cal {T}}(E_{\gamma})}{ E_{\gamma}^3 },
\end{equation}
as dipole transitions dominate the decay in the quasi-continuum \cite{kopecky1990}.
We propagate the uncertainties from both the normalization of the NLD and the $\gamma$SF to obtain a total uncertainty band as shown in Fig.~\ref{fig:gsfEB}.
Also shown in the figure are the separate contributions bands from the NLD and the $\gamma$Sf normalization. 
It is clear that the contribution from the NLD (i.e. the $\rho(S_n)$ estimate) is much smaller than the uncertainty from the scaling parameter $B$, i.e. the $\left< \Gamma_{\gamma0}\right>$ value dominates the total uncertainty.  

\section{Shell-model calculations and comparison to auxiliary NLD data}
\label{sec:KSHELL}

To investigate the NLD of $^{64}$Zn in more detail, we have performed calculations within the configuration-interaction shell model using the \textsf{KSHELL} code~\cite{KSHELL}, which is based on the $m$-scheme basis.
We have applied the {\sf gs8} interaction~\cite{gs8_interaction}, which comprises the \textit{sd-pf-sdg} single-particle orbitals with $^{16}$O as the inert core. 
Due to the large model space it was necessary to apply a $1\hbar \omega$ truncation, and we have also required a minimum occupation of 14 nucleons in the two $1f_{7/2}$  orbitals (protons and neutrons combined).
With these truncations, the $m$-scheme dimension becomes $D_m \approx 1.59 \times 10^8$.
For the spins, we have set a limit of maximum 200 levels per spin for $J = 0, 1, \dots, 16$, and for both parities. 
We have run the calculations on the supercomputer ``Betzy''\footnote{\url{https://documentation.sigma2.no/hpc_machines/betzy.html}} maintained by the Norwegian National e-infrastructure services \textsf{Sigma2}. 

\begin{figure}[t]
    \includegraphics[clip,width=1.0\columnwidth]{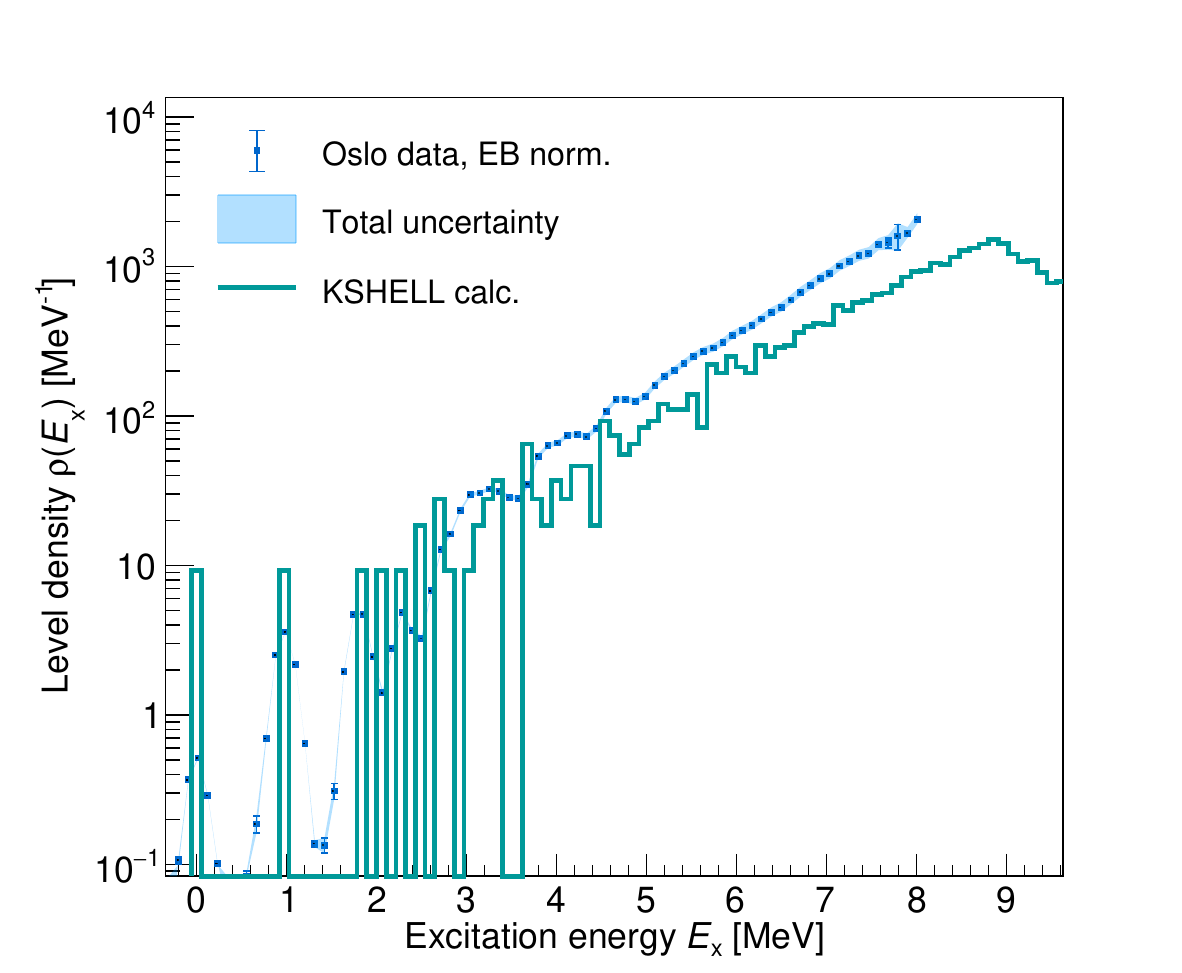}
    \caption{(Color online) Extracted level density (filled data points) of $^{64}$Zn with the EB normalization. 
    The histogram shows the  shell-model results, with the same bin width as the experimental data points (108 keV).}
    \label{fig:countingx}
\end{figure}

The resulting NLD for all the included spins and both parities is compared to the Oslo results and the discrete levels in Fig.~\ref{fig:countingx}.
Although the calculated NLD is lower in absolute value than the experimental data, the calculations describe very well the known discrete levels as well as the constant-temperature like behavior seen in our data for excitation energies between $E_x \approx 4-8$ MeV. 
At $E_x \approx 8.8$ MeV, the levels with the most common spins ($J=2-6$) are already spent, as we have a maximum of 200 levels per spin and parity in the calculations. 
Therefore, at higher $E_x$, the calculated level density is no longer complete and we see a drop in the \textsf{KSHELL} NLD. 

\begin{figure}[t]
    \includegraphics[clip,width=0.8\columnwidth]{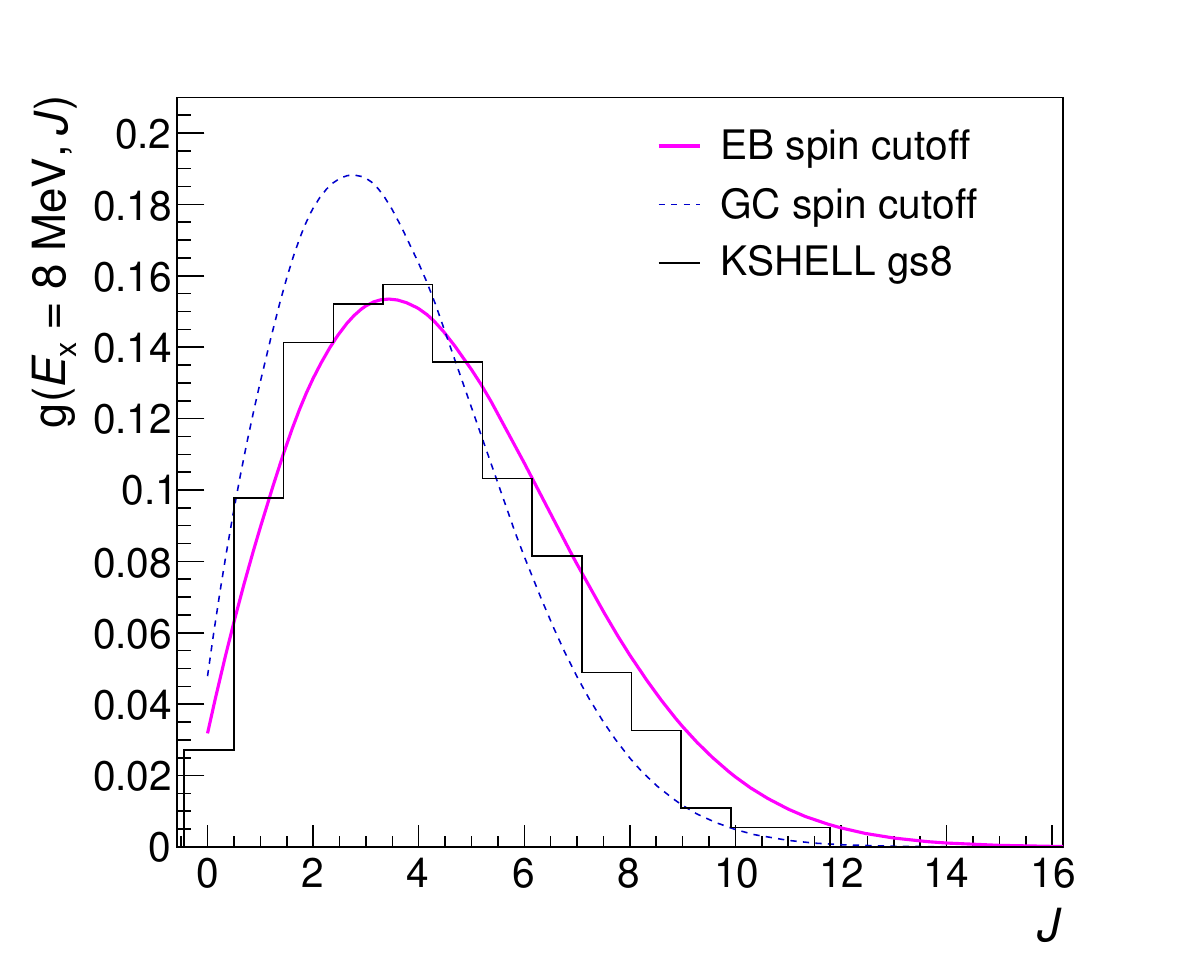}
    \caption{(Color online) Calculated spin distribution of $^{64}$Zn at $E_x = 8.0$ MeV with the EB (solid, magenta line) and GC (blue dashed line) parameterizations, respectively. 
    The histogram shows the shell-model result.}
    \label{fig:spindist}
\end{figure}

In Fig.~\ref{fig:spindist},  we show the calculated spin distribution for $^{64}$Zn at $E_x = 8.0$ MeV and compare with the spin distribution given in Eq.~\ref{eq:spindist} with the EB and GC phenomenological spin cutoff parameters.
Overall, we find that the shell-model calculations are similar to the EB model which assumes a rigid-body moment of inertia. 
Thus the shell-model calculations provide support for the EB model being the most appropriate to use for $^{64}$Zn. 

\begin{figure}[bt]
    \includegraphics[clip,width=1.0\columnwidth]{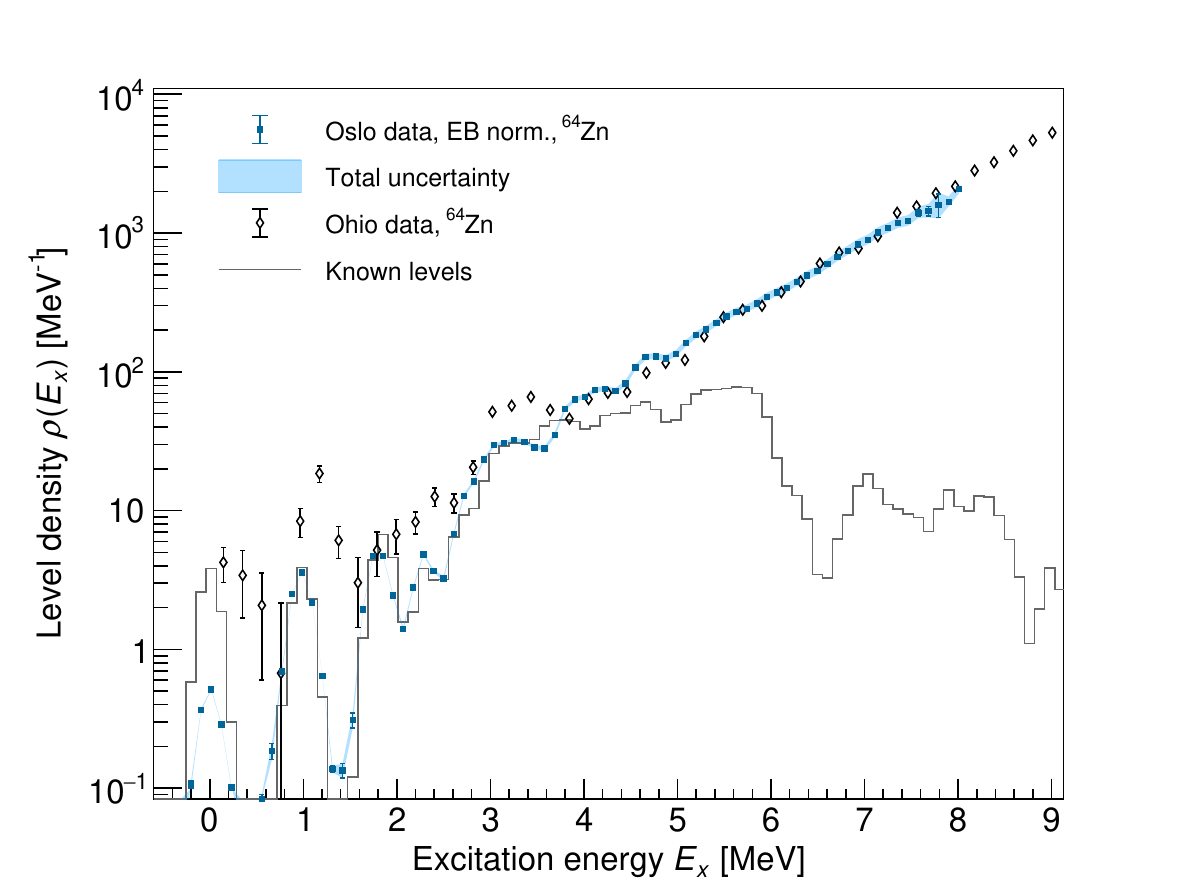}
    \caption{(Color online) The Oslo level density (filled data points) of $^{64}$Zn with the EB normalization compared to Ohio data applying the particle-evaporation method~\cite{Ramirez2013} (open diamonds). The Ohio data have been re-scaled with a factor of 0.9 (see text). 
    The histogram shows the binned, discrete levels.}
    \label{fig:nld_with_Ohio}
\end{figure}

In Fig.~\ref{fig:nld_with_Ohio}, we show a comparison of our data with previous measurements performed by the Ohio group using the particle-evaporation method on data from the $^{63}$Cu($d,n$)$^{64}$Zn reaction~\cite{Ramirez2013}.
Overall, we observe a good agreement between the two methods, although the Ohio data show a significant overshoot with respect to both the Oslo data and the discrete levels for $E_x < 4.0$ MeV. 
However, as discussed in Ref.~\cite{Ramirez2013}, the presence of non-compound reaction mechanisms for high-energy neutrons (corresponding to low $E_x$ in the residual nucleus) from the ($d,n$) reaction could lead to enhancements in the differential cross sections which cannot be described by a pure Hauser-Feshbach-type model.
This makes the normalization of the Ohio data to the discrete levels more difficult and could lead to a sizable uncertainty in the absolute normalization.
To provide a more useful comparison between the data, we have therefore performed a slight re-scaling of the Ohio data with a factor of 0.9~\cite{Voinov2025}. 
Also, new, preliminary Ohio data from the $^{58}$Ni($^7$Li,$p$)$^{64}$Zn reaction confirm that the Oslo normalization is reasonable~\cite{Voinov2025}.
We further note that our data support the findings of Ref.~\cite{Ramirez2013} in that the CT model fits best with the data at high excitation energies, and we obtain temperature parameters of $T_{\mathrm{GC}} = 1.20(2)$ MeV and $T_{\mathrm{EB}} = 1.17(2)$ MeV, in very good agreement with $T_{\mathrm{Ohio}} = 1.19(2)$ MeV deduced from the data for a deuteron beam energy of 7.5 MeV.

\section{The shape method and non-statistical decay to the ground state}
\label{sec:shapemethod}
\begin{figure*}[tb]
\begin{center}
\includegraphics[clip,width=2\columnwidth]{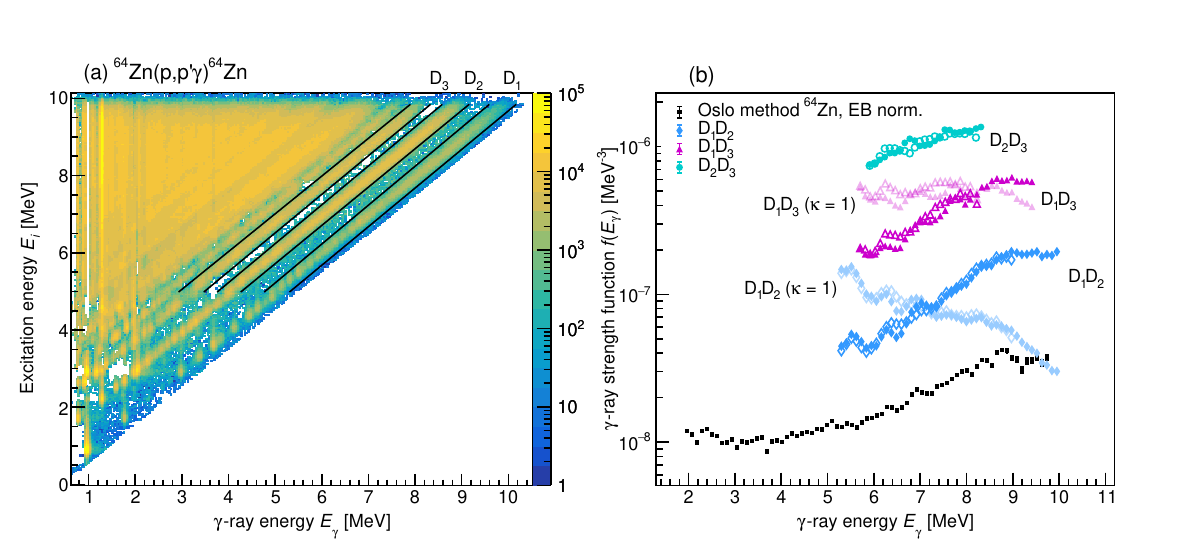}
\caption{(Color online) (a) The primary $\gamma$ matrix $P(E_{\gamma}, E_i)$ of $^{64}$Zn showing the cuts for the three diagonals. The diagonals $D_1$, $D_2$ and $D_3$ contain the number of transitions of direct decay from the quasi-continuum feeding into the $0^+$ ground state, the first excited $2^+$ level (992 keV) and the second $2^+$ and $0^+$ levels (1799 and 1910 keV), respectively. (b) The $\gamma$SFs from the shape method (filled and open colored markers) are scaled for better visualization. The data points with transparent markers show the results without a hindrance factor ($\kappa=1$).
The applied hindrance factor for the $D_1D_2$ and $D_1D_3$ data points is $\kappa=0.52$ (see Appendix~\ref{appB}).
The shape-method data points include only statistical uncertainties. 
}
    \label{fig:shape}
\end{center}
\end{figure*}

Figure~\ref{fig:shape} (a) shows the primary $\gamma$ matrix displaying ``diagonals'' $D_1$, $D_2$ and $D_3$, which represent direct decay from the quasi-continuum to specific, discrete final levels. 
We will now apply the shape method presented in Wiedeking \textit{et al.}~\cite{Wiedeking2021} using these diagonals. 
The extracted shape-method data are evaluated with the code {\sf diablo.c} available on the Oslo Cyclotron GitHub~\cite{om2022}. 

The main idea behind the shape method is to measure the number of counts $N_D$ within a diagonal $D$  at a given initial excitation-energy bin $E_i=E_{\gamma}+E_f$ for a fixed final excitation energy $E_f$. 
For $D_1$, the final level is only the ground state with $J_f = 0^+$.
For $D_2$, the first excited $2^+$ is the final level, while for $D_3$, the second  $2^+$ and  $0^+$ are the final levels. 
Under the assumption that we have an equal amount of levels with positive and negative parity for the initially populated levels, the number of $\gamma$-ray transitions with energy $E_{\gamma}$ from a given initial excitation energy $E_i$, obeys the following proportionality:
\begin{equation}
 N_D(E_i,E_f,J_f)\propto f(E_{\gamma})E_{\gamma}^3 \sum_{[J_f]}\sum_{J_i=J_f-1} ^{J_i=J_f+1}p^{\rm level }(E_i,J_i) \;g (E_i,J_i),
 \label{eq:ndshape}
\end{equation}
where $p^{\rm level}(E_i,J_i)$ denotes the probability for populating levels at $E_i$ with a given spin $J_i$, and $g (E_i,J_i)$ is the spin distribution given in Eq.~(\ref{eq:spindist}). 
In the following, we make the assumption that the population cross-section in the quasi-continuum is approximately independent of spin, i.e.~$p^{\rm level }$ is just a constant and the same for all $E_i$ and $J_i$. 
All transitions are assumed to be of dipole type as the dipole strength is known to be dominant within the quasi-continuum~\cite{kopecky1990}. The notation $[J_f]$ indicates the (set of the) spin(s) of the final level(s) within a diagonal. 
The second sum is restricted to the available initial spins $J_i$ given by the final spin(s) $J_f$ and the restriction of dipole transitions ($M1$ or $E1$). For the final level being the ground state with $J_f=0$, only initial spins $J_i=1$  are allowed. 
 For $J_f>0$, in general three initial spins are available; \textit{e.g.} for $J_f=2$, $J_i = 1,2,3$. 

By applying gates on the diagonals as shown in Fig.~\ref{fig:shape}~(b), we can now construct pairs of intensities $(N_{D1},N_{D2})$, $(N_{D1},N_{D3})$ and $(N_{D2},N_{D3})$ which, when corrected for the different initial spin distribution in the case where the $J_f$ values are different, should reflect the shape of the $\gamma$SF when the pairs are internally normalized for each $E_i$ bin by a sewing technique based on a logarithmic interpolation. 
In the following, the strength obtained with these pairs are called $D_1D_2$, $D_1D_3$ and $D_2D_3$ respectively.
The obtained strength function $f(E_{\gamma})$ has in principle the correct functional form, although the absolute normalization is arbitrary.

The shape method results are shown in Fig.~\ref{fig:shape} (b), where filled, colored markers represent the decay to the lowest $E_f$ while open colored data points represent decay to the upper $E_f$. 
We see immediately that both the $D_1D_2$ and $D_1D_3$ combinations have a very unexpected behavior, with a strength that is seemingly decreasing as the $\gamma$-ray energy increases. 
To investigate the cause of this behavior, we have made a closer examination of the Oslo-method level density of the levels at the $E_f$ bins for the $D_1$ and $D_2$ gates. 

\begin{figure}[t]
    \includegraphics[clip,width=1.0\columnwidth]{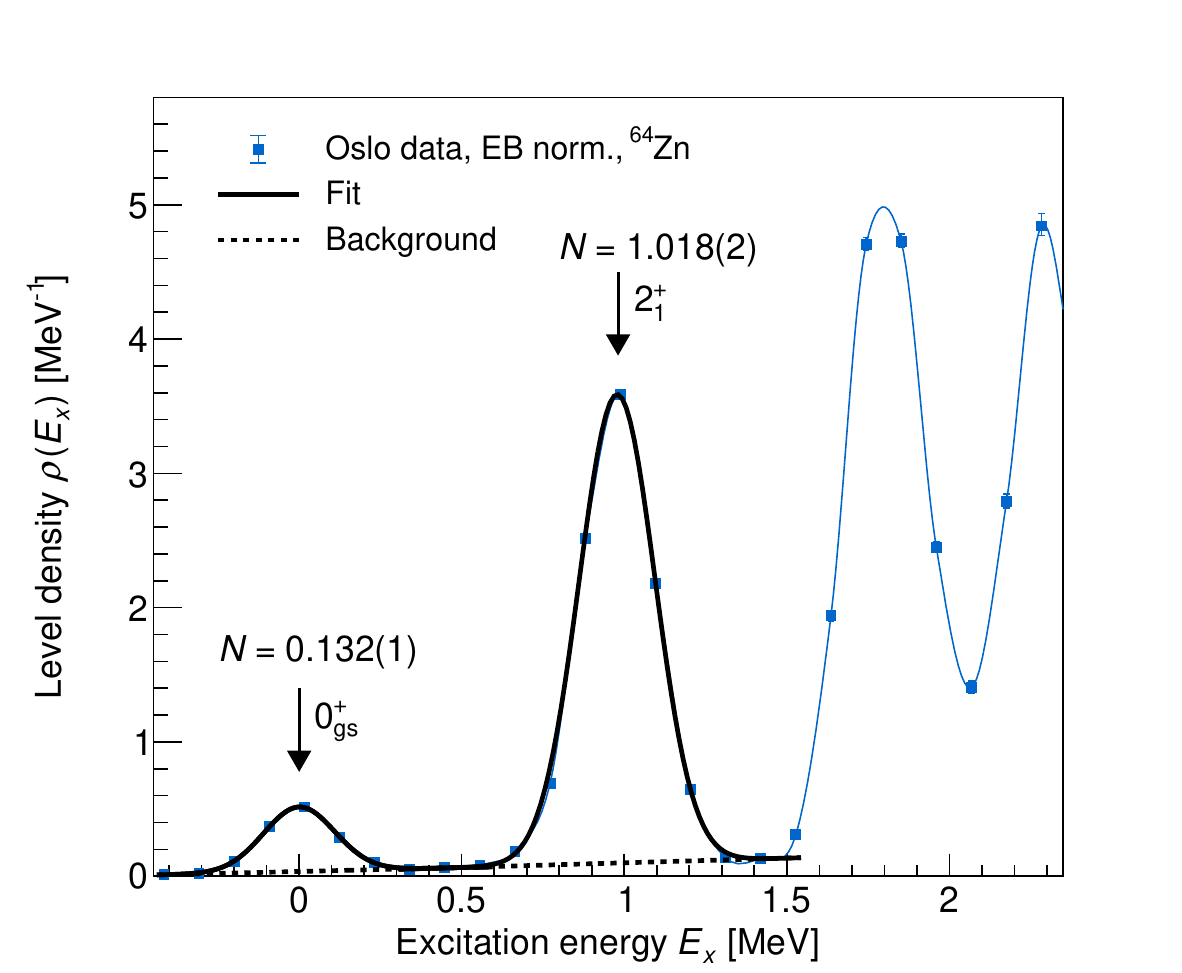}
    \caption{(Color online) The  level density of $^{64}$Zn in the low excitation-energy region on a linear scale (filled data points) for the EB normalization. The spin and parity for the ground state and first excited level are indicated in the figure. The  number of levels $N$   
    is determined by integrating the Gaussian fit to the peaks. }
    \label{fig:countingx_lin}
\end{figure}

A striking feature in Fig.~\ref{fig:counting_EB} is that the level density of the ground state appears to be much lower than the first $2^+$ level in the Oslo data. 
As mentioned previously, only initial levels with $1^+$ and $1^-$ decay directly to the ground state, whereas the $2^+$ level is reached by $1^+$, $1^-$, $2^+$, $2^-$, $3^+$  and $3^-$ levels through dipole decay.
If the intrinsic spin distribution was flat, one could expect an intensity reduction of 1/3 for the $0^+$  level density.
Here, we use the average initial excitation energy  for $E_i =$ 4.98--9.84 MeV (the $E_i$ range applied in the shape method), $\left<E_i\right> = 7.41$ MeV, obtaining the spin-cutoff value from Eq.~(\ref{eq:sigE}): $\left<\sigma_J\right> = 3.53$.
From this, we deduce from the $g(\left<E_i\right>,J)$ distribution an average intensity-reduction factor: 
${g(J=1)}/\left[g(J=1)+g(J=2)+g(J=3)\right] = 0.251$. 

From Fig.~\ref{fig:countingx_lin},
it is clearly seen that the ground-state level density is significantly lower than the $2^+_1$ level density. 
From a fit to the ground state and the  $2^+_1$ level assuming a simple, linear background, we find that the ground state is  $\approx 13$\% of a full level, while the  $2^+_1$ level integrates to approximately one level as it is expected to be. 
For the GC normalization we obtain similar results. 
It therefore seems like the decay to the ground state is strongly hindered, as the suppression of the level density is much bigger than the factor of $\approx 0.25$ expected from the spin distribution of the initial levels alone. 
In Appendix~\ref{appB}, we test various hypotheses and conclude that the integrated level densities of the ground state $0^+_\text{gs}$, the first excited level $2_1^+$, and the second $0^+_\text{2},2_2^+$ are consistent with a hindrance factor of $\kappa\approx 0.5$ for $\gamma$ decay to the ground state. 
Such a hindrance factor seems either not to be present, or present to a lesser degree, for the decay to the second $0^+_2$ level.

One could wonder why the direct decay to the ground state is significantly suppressed. 
It is tempting to search for an explanation related to the different structures of the three lowest $0^+$ levels. 
If the wave functions of the quasi-continuum levels have a larger overlap with the wave function of the second $0^+$ level, more decay strength could be funneled through this level. 
This could also be a sign of significantly different nuclear shapes of the ground state vs. the excited states, as discussed in the Introduction. However, at present this is only speculation.

Figure~\ref{fig:shape} (b) shows that by assuming a hindered decay to the $0^+$ ground level with a hindrance factor $\kappa \approx 0.5$ (see hypothesis H1 of Appendix~\ref{appB}), all three $\gamma$SFs give consistent slopes with the Oslo method result. 
In practice, the hindrance factor is implemented by replacing the counts $N_{D1}$ of the $0^+_\text{gs}$ diagonal with $N_{D1}/\kappa$. As expected, if we assume $\kappa=1$, the $D_1D_2$ and $D_1D_3$ combinations fail completely to predict a realistic strength function (transparent markers). 
Equation (\ref{eq:array3_1}) shows that $D_2D_3$ is independent of $\kappa$, which means that this strength function remains unchanged.
The fact that all three diagonal combinations (when applying $\kappa \approx 0.5$) display approximately the same slope as the Oslo-method $\gamma$SF, provides further support for the $\rho(S_n)$ value obtained from the systematics.

\section{Comparison to other $\gamma$SF measurements and models of the dipole $\gamma$SF}

\begin{figure}[t]
\begin{center}
\includegraphics[clip,width=1.\columnwidth]{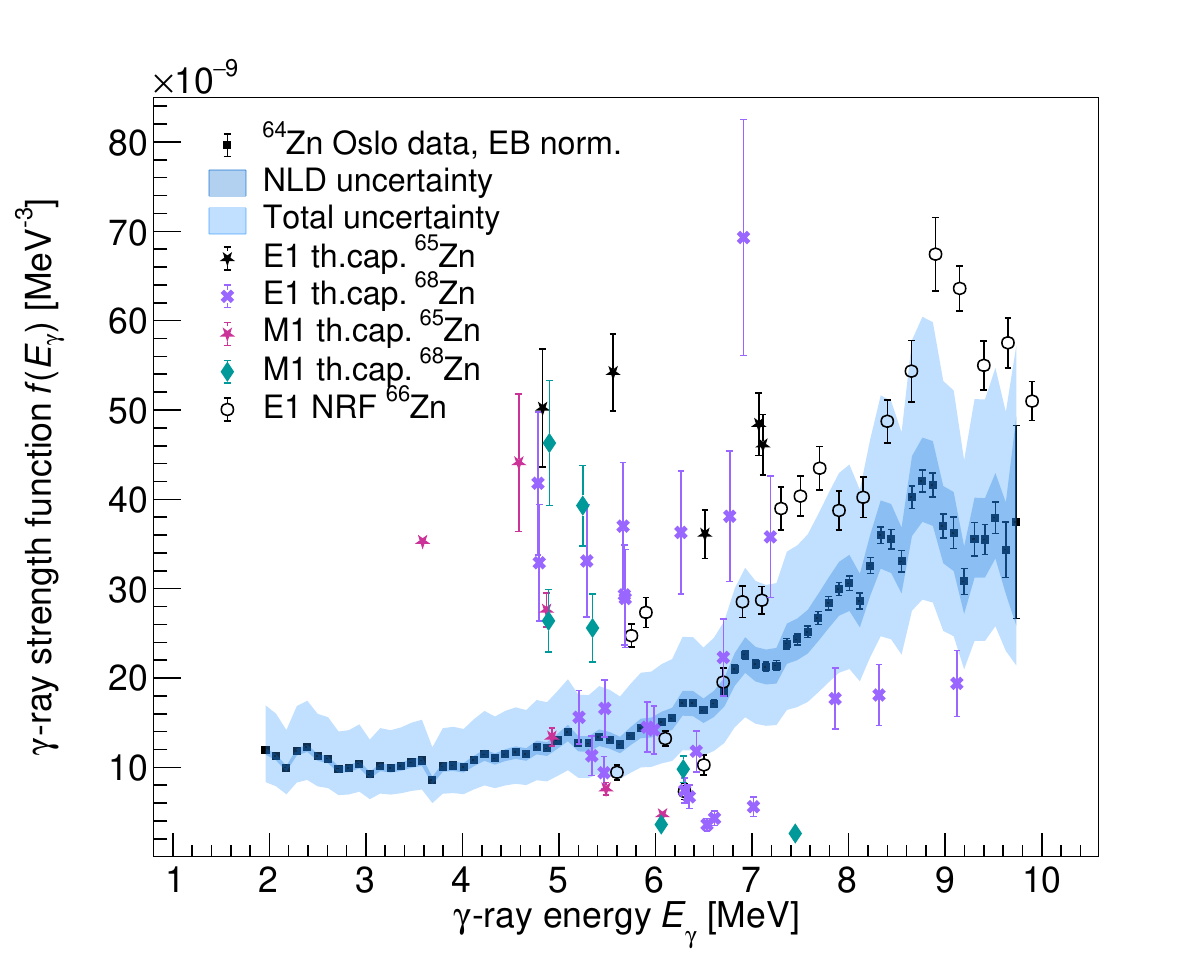}
\caption{(Color online) The $\gamma$SF results from the Oslo data compared to $E1$ and $M1$ strengths extracted from primary transitions following radiative neutron capture in $^{65,68}$Zn~\cite{IAEA2022,Goriely2019} and NRF data from the HI$\gamma$S facility for $^{66}$Zn~\cite{savran2022}. }
    \label{fig:gsf_other_data}
\end{center}
\end{figure}

We now compare our $^{64}$Zn $\gamma$SF with data of other Zn isotopes below the neutron threshold, see Fig.~\ref{fig:gsf_other_data}.
We observe a very large spread in the individual data points for both the $E1$ and $M1$ strengths derived from thermal neutron capture, with a ratio of $\approx 19$ and $\approx 18$, respectively, of the largest and smallest values.
In the Nuclear Resonance Fluorescence (NRF) experiment of Savran~{\em et al.}~\cite{savran2022}  photoabsorption cross section data for $^{66}$Zn were obtained using monoenergetic photon beams at the High Intensity $\gamma$-ray Source (HI$\gamma$S) facility.   
The $\gamma$SF of Savran~{\em et al.}~\cite{savran2022} (open data points) reveals stronger fluctuations below $E_{\gamma}\approx 8$~MeV; however, the data points above $E_{\gamma}\approx 8$~MeV are in reasonable agreement with the upper part of the $1\sigma$ uncertainty band estimated for $^{64}$Zn. 
In addition, the overall structures of the NRF data agree well with the Oslo method data.

\begin{figure}[t]
\begin{center}
\includegraphics[clip,width=1.\columnwidth]{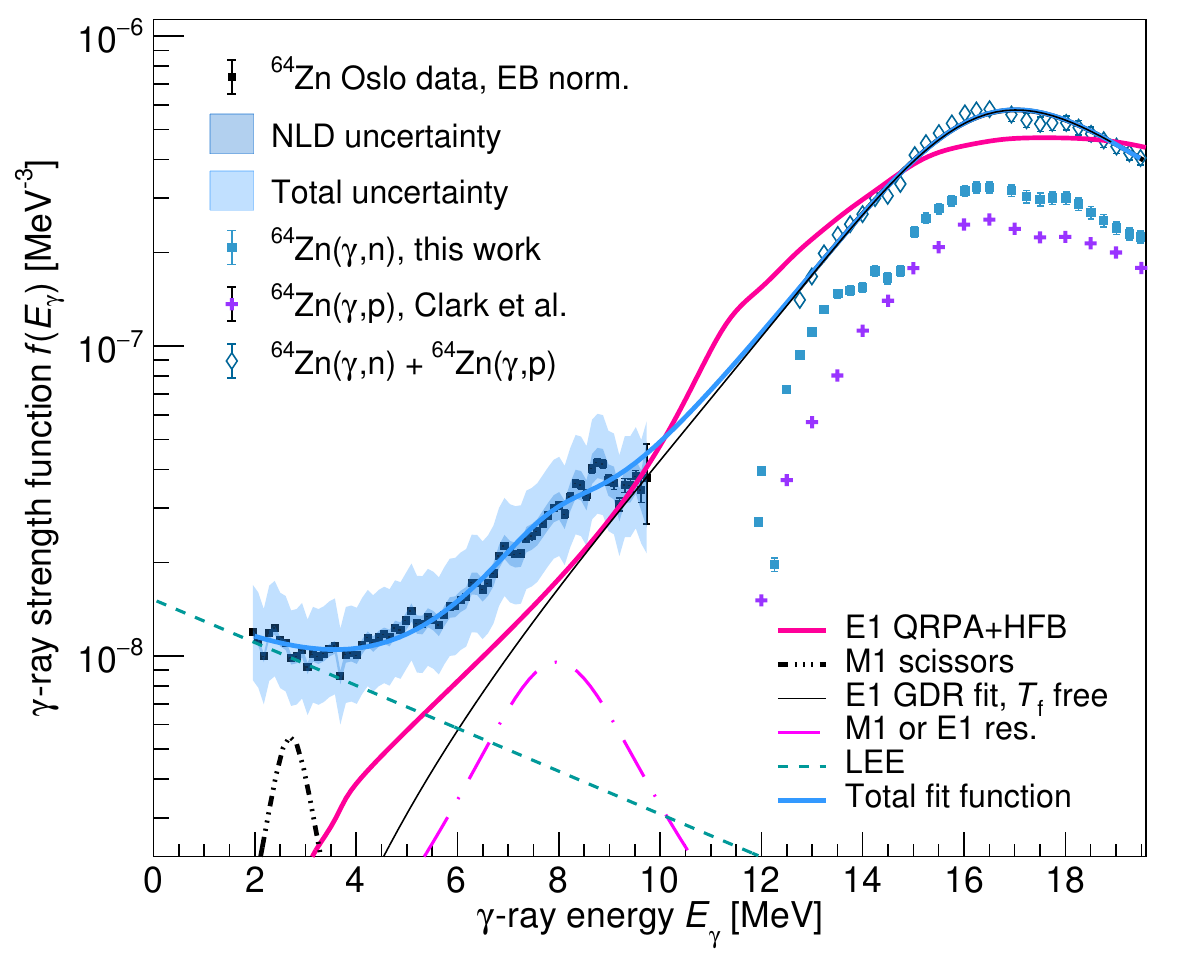}
\caption{(Color online) Comparison with $E1$ QRPA+HFB calculations~\cite{Goriely2004} (magenta line) and fits to the $^{64}$Zn $\gamma$SF Oslo data, the NewSUBARU $^{64}$Zn$(\gamma,n)$  data and the $^{64}$Zn$(\gamma,p)$ data from Clark {\em et al.}~\cite{clark1973} (see text). Here, the $T_f$ parameter in the GLO model has been treated as a free parameter with a range of $T_f \in [0.0,1.5]$ MeV. }
    \label{fig:totalgsf}
\end{center}
\end{figure}

We now proceed to fit our Oslo-method data and the photo-absorption data to semi-empirical models for the $\gamma$SF, as shown in Fig.~\ref{fig:totalgsf}. 
  
\begin{figure}[t]
\begin{center}
\includegraphics[clip,width=1.\columnwidth]{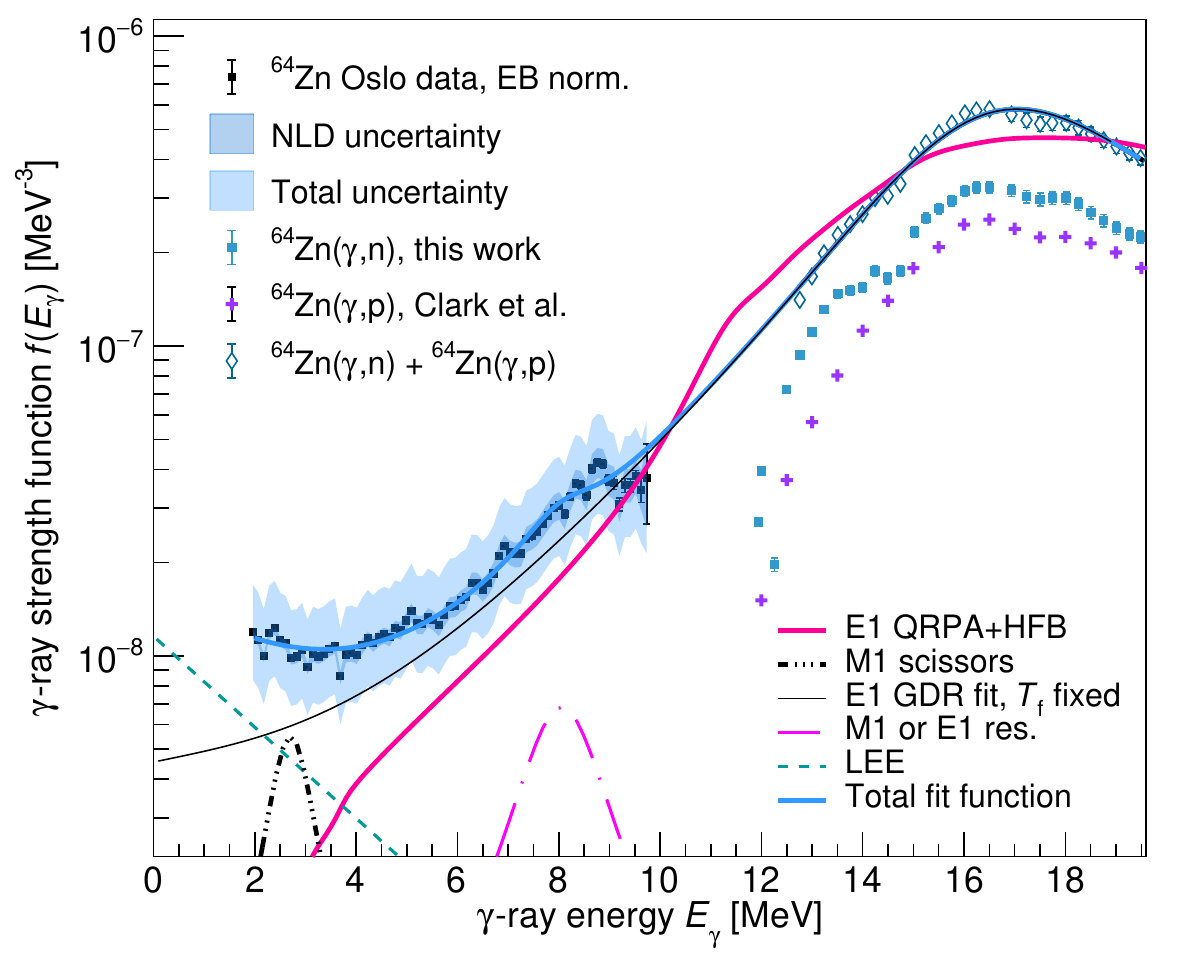}
\caption{(Color online) Same as Fig-~\ref{fig:totalgsf}, but with the $T_f$ parameter in the GLO model fixed to a value of $T_f = 0.7$ MeV. }
    \label{fig:totalgsf_fixed_Tf}
\end{center}
\end{figure}

For the $\gamma$SF in the region above $E_{\gamma}=S_n$, the ($\gamma, n$) and ($\gamma, p$) cross sections are transformed to $\gamma$SF by means of the relation~\cite{Axel1968}
\begin{equation}
f(E_{\gamma})= \frac{1}{3(\pi \hbar c)^2}\frac{\sigma_{\gamma x}(E_{\gamma})}{E_{\gamma}},
 \label{eq:gsf}
\end{equation}
where the constant reads $1/3(\pi \hbar c)^2=8.674\times 10^{-8}$mb$^{-1}$MeV$^{-2}$, and $x$ is either $n$ (neutron) or $p$ (proton). 
In this case, since the proton separation energy is quite low for $^{64}$Zn ($S_p=7.713$~MeV), the $(\gamma,p)$ cross-section data of Clark {\em et al}.~\cite{clark1973} contribute significantly to the total strength.
Therefore, for the comparison with microscopic calculations, as well as input data for the fit to empirical models, we have added the two datasets. The resulting total $\gamma$SFs of $(\gamma,n)$ and $(\gamma,p)$ data are shown as open diamonds in Figs.~\ref{fig:totalgsf} and~\ref{fig:totalgsf_fixed_Tf}.
The total experimental $\gamma$SF above $S_n$ is clearly dominated by the giant dipole resonance (GDR) centered between $E_{\gamma}\approx 14.8-18.5$~MeV. 
In the ($\gamma, n$) cross-section data, in addition to the maxima at $E_\gamma \approx 16.3$ MeV and $\approx 18.0$ MeV, a broad structure appears at $E_{\gamma}\approx 13.5$~MeV, for which the origin is unclear. 
The structure might be due to an $E1$ fragmentation with a weak degree of collectivity, as discussed by Reinhard and Nazarewicz~\cite{reinhard2013}.
On the other hand, the three bumps in the ($\gamma, n$) data might also be an indication of the ground state of $^{64}$Zn being triaxial, as discussed in the Introduction.
We note that the microscopic quasiparticle random-phase approximation (QRPA) plus Hartree-Fock-Bogoliubov (HFB) calculations of Ref.~\cite{Goriely2004} capture more or less the general behavior for the photo-cross section data, although there is some mismatch of the absolute value at the peak of the GDR as well as the width.
 
For the fit of the total experimental $\gamma$SF, we assume the composite function
\begin{equation}
     f(E_{\gamma})= f_{\rm GDR}+f_{\rm PDR}+f_{\rm LEE},
     \label{eq:totalgsf}
\end{equation}
where the various terms are described below.

\begin{table*}[t]
\caption{Fitted GDR, PDR and LEE parameters to the experimental $\gamma$SF of $^{64}$Zn.
The first line gives the resulting parameters when $T_f$ is a free parameter, while in the second line, $T_f$ is fixed.}
\begin{tabular}{ccccccccc}
\hline
\hline
\multicolumn{4}{c}{GDR}  &\multicolumn{3}{c}{PDR}&\multicolumn{2}{c}{LEE}\\
\cmidrule(rl){1-4} \cmidrule(rl){5-7} \cmidrule(rl){8-9} 
 $E$ &$\sigma$  &$\Gamma$ & $T_f$ &$E$  &$\sigma$ &$\Gamma$ &   $C$    &$\eta$      \\
 $[$MeV$]$ &   [mb]   &  [MeV] & [MeV]  & [MeV] & [mb]  &  [MeV] & [$10^{-8}$MeV$^{-3}$] &[MeV$^{-1}$]\\
\hline
18.2(1)&  109(2)&  9.1(4) & 0.0(14) &8.1(5)& 0.88(36)&   3.0(36) & 1.5(5) &0.16(15)  \\

18.1(1)&  116(2)&  8.1(3) & 0.7 &8.1(6)& 0.64(48)&   1.8(16) & 1.6(8) &0.34(24)  \\ 
\hline
\hline
\end{tabular}
\\
\label{tab:GDR}
\end{table*}


To describe the GDR data, we use the generalized Lorentzian (GLO) with the functional form of~\cite{kopecky1990}:
\begin{align}
&f_{\rm GDR}(E_{\gamma}) = \frac{1}{3\pi^2\hbar^2c^2}\sigma\Gamma   \nonumber \\
                            & \times \left[ \frac{ E_{\gamma} \Gamma(E_{\gamma},T_f)}{(E_\gamma^2-E^2)^2 + E_{\gamma}^2 \Gamma^2(E_{\gamma},T_f)}
                             + 0.7 \frac{\Gamma(E_{\gamma}=0,T_f)}{E^3}\right]
\label{eq:GLO}
\end{align}
with
\begin{equation}
\Gamma(E_{\gamma},T_f) = \frac{\Gamma}{E^2} (E_{\gamma}^2 + 4\pi^2 T_f^2).
\end{equation}
The resonance parameters are the energy centroid $E$, the strength $\sigma$, and the width $\Gamma$. The $T_f$ parameter, a ``temperature'' of the final states in the de-excitation, can provide a non-zero tail for the lower $\gamma$ energies. 
In principle, we could have fitted three functions to get a better description of the broad structures at $E_\gamma \approx 13.5,\, 16.3, \, 18.0$ MeV in the summed ($\gamma,n$)+($\gamma,p$) data. 
To avoid having a large number of fit parameters, we choose here to use a single GLO curve.

The pygmy dipole resonance (PDR) is traditionally believed to originate from oscillations of the neutron skin against the $N=Z=30$ core. The number of excess neutrons $\delta N = A-2Z=4$ is very small for $^{64}$Zn, which puts such an interpretation in question.  
Moreover, proton-scattering experiments by Djalali \textit{et al.}~\cite{Djalali1982} have revealed $M1$-type resonance-like structures between $E_\gamma \approx 8.0-10$ MeV in $^{68}$Zn. 
We therefore add into the fit function a resonance described by the standard Lorentzian (SLO) model~\cite{brink,Axel1962} given by
\begin{equation}
f_{\rm PDR}(E_{\gamma}) = \frac{1}{3\pi^2\hbar^2c^2}\frac{\sigma E_{\gamma} \Gamma^2}{(E_\gamma^2-E^2)^2 + E_{\gamma}^2 \Gamma^2},
 \label{eq:SLO}
\end{equation}
with the centroid $E$, peak cross section $\sigma$, and width $\Gamma$ as fit parameters.
It is listed as ``PDR'' in Eq.~(\ref{eq:totalgsf}) and in Table~\ref{tab:GDR}, but as the Oslo method does not enable us to distinguish between $E1$ and $M1$ transitions, this structure could either be of $E1$ or $M1$ type, or a combination of the two.

From numerous shell-model calculations (\textit{e.g.} Refs.~\cite{schwengner2013,Brown2014,Sieja2017,Midtbo2018,Dahl2026}), and the projected shell model calculations~\cite{Chen2025}, the low-energy enhancement (LEE) has been predicted to be due to magnetic dipole transitions.
The $M1$ nature of the LEE has recently been confirmed experimentally by Ronning \textit{et al.} in $^{70}$Zn~\cite{Ronning2026}.
An exponential form of the LEE was suggested by Schwengner {\em et al.}~\cite{schwengner2013}:
\begin{equation}
f_{\rm LEE}(E_{\gamma}) = C \exp (-\eta E_{\gamma}),
 \label{eq:LEE}
\end{equation}
where $C$ gives the absolute value and $\eta$ is the slope. 

Figure~\ref{fig:totalgsf} shows the obtained fit function of Eq.~(\ref{eq:totalgsf}) (blue curve), which includes nine parameters. 
The fit method of the ROOT data analysis package\footnote{\href{https://root.cern}{ROOT 6.36.04  @ CERN}} has been applied, which is based on the Minuit package~\cite{james1981} with Hessian matrix error analysis. 
Here, we take two different approaches to determining $T_f$, which strongly influences the modeled $E1$ GDR tail. 
First, we allow this parameter to be a free parameter obtained from the fit to the data, within the range $T_f \in [0.0,1.5]$ MeV.
The resulting fit is shown in Fig.~\ref{fig:totalgsf}.
Next, we fix this parameter to $T_f = 0.7$ MeV, 
which significantly increases the $E1$ GDR tail, as seen in Fig.~\ref{fig:totalgsf_fixed_Tf}.
This value, $T_f = 0.7$ MeV, was chosen so that the $E1$ tail increases significantly in comparison with the QRPA prediction and also when $T_f$ is free, but at the same time allowing for some $M1$ component of the $\gamma$SF, in accordance with external data for other zinc isotopes (see Fig.~\ref{fig:gsf_other_data}.
The resulting parameters  are listed in Table~\ref{tab:GDR}. 

We note that some of the obtained fit parameters in Table~\ref{tab:GDR} have very large uncertainties, this concerns in particular the peak cross section and width of the PDR, and the slope parameter $\eta$ of the LEE. 
This is mainly caused by the uncertain $T_f$ parameter as well as the uncertainty in the estimated $\left< \Gamma_{\gamma0}\right>$. 
If we were to use the Oslo-method data with just the uncertainties stemming from the Oslo method itself, or including only the NLD uncertainties in the fit, we do obtain much smaller uncertainties especially for the LEE slope. 
As $\left< \Gamma_{\gamma0}\right>$ gives the absolute value for the whole $\gamma$SF, i.e., it determines the common scaling factor $B$ for all the individual data points, it does not change the trend of the data points which has a minimum at $E_\gamma \approx 4$ MeV and then increases for lower $E_\gamma$.
Thus, we can conclude that the LEE is present in the $\gamma$SF of $^{64}$Zn, similar to the findings in $^{70}$Zn from beta-decay studies~\cite{Ronning2026}.

We remark that our $\gamma$SF data do not show any clear sign of an $M1$ scissors mode, which would be expected to manifest itself as a pronounced, broad bump centered around $E_\gamma \approx 2.7$~MeV. Assuming a deformation parameter $|\delta| = 0.22$ and making use of the sum-rule approach by Enders \textit{et al.}~\cite{Enders2005} with a rigid-body moment of inertia as in Ref.~\cite{Guttormsen2014}, the expected integrated strength would yield $\sum B(M1) \approx 2 \mu_N^2$.
If we further assume that the scissors mode can be described with a Standard Lorentzian~\cite{brink} (see Eq.~(\ref{eq:SLO})), with a width $\Gamma \approx 1.0$ MeV and peak cross section $\sigma \approx 0.17$ mb (in accordance with the integrated sum-rule strength), we see in Figs.~\ref{fig:totalgsf} and~\ref{fig:totalgsf_fixed_Tf} that the contribution of a possible scissors mode remains  small.
Thus, it is possible that it is present, but with a strength too weak for us to be able to identify it in the experimental $\gamma$SF of $^{64}$Zn.

\section{Summary and outlook}
In this work, we have presented new results on the NLD and $\gamma$SF of $^{64}$Zn from  $(p, p' \gamma)$ data measured at the Oslo Cyclotron Laboratory and $(\gamma, n)$ data measured at NewSUBARU. 

The NLD of $^{64}$Zn is extracted from the ground state up to $\approx 8 $~MeV of excitation energy, and displays an overall constant-temperature-like shape, in agreement with previously measured particle-evaporation data and large-scale shell model calculations presented in this work. 
The observed low level density of the ground state indicates that the decay to this level is strongly hindered. By introducing a hindrance factor $\kappa \approx 0.5$, the shape method gives consistent slopes in the extracted $\gamma$SFs for all three diagonals of the primary matrix.
This interesting behaviour is also observed in an ongoing analysis of Oslo-method data for $^{70}$Zn.

By combining the Oslo data and the NewSUBARU data together with previously measured ($\gamma,p$) cross sections, the experimental $\gamma$SF covers the main part of the $E_{\gamma} \approx$2--19~MeV energy region. 
The lower energy part of the $\gamma$SF is mainly described by the tail of the GDR. 
Superimposed on the $E1$ tail of the GDR, our data are compatible with a significant LEE structure for $E_\gamma < 4$ MeV.
To further constrain the parameters of the LEE would require better constraints on the $E1$ component and the average total radiative width.
We find no clear evidence for an $M1$ scissors mode in the data, however we cannot exclude the presence of a weak $M1$ scissors around $E_\gamma \approx 2.7$ MeV. 

In future works, it would be highly desirable to further explore the degree of hindrance in the decay to the ground state in other Zn isotopes and other nuclei in this mass region. Moreover, it could be that the $M1$ scissors mode is more pronounced in heavier Zn isotopes; this remains to be investigated.

\begin{acknowledgments}
We  thank J.~C.~M\"uller, P.~Sobas and J.~Wikne for providing excellent  experimental conditions.
We are grateful for the help of K.~S.~Beckmann on the OCL experiment. 
A.~C.~L. would like to sincerely thank A.~V.~Voinov and A.~P.~D.~Ramirez for their help with the Ohio data as well as for many fruitful and stimulating discussions.
The \textsf{KSHELL} calculations were performed on resources provided by Sigma2, the National Infrastructure for High Performance Computing and Data Storage in Norway using ``Betzy'' on Project No. NN9464K.
J.~K.~D. and A.~C.~L. gratefully acknowledge continued support from the Centre for Computational and Data Science (dScience) at the University of Oslo, Norway.
A. C. L. gratefully acknowledges funding
from ERC-STG-2014 under Grant Agreement No. 637686
the Research Council of Norway, Project No. 316116.
This work was also partially supported by the Research Council of Norway: Projects No. 263030, 262952,
325714 and the Norwegian Nuclear Research Centre (Project No. 341985). 
The OSCAR detector was funded by the Research Council of Norway, Project No. 245882.
This material is based upon work supported by the U.S. Department of Energy, Office of Science, Office of Nuclear Physics under Contract No. DE-AC02-05CH11231, the U.S. Nuclear Data Program and by the National Research Foundation of South Africa (Grant Number: 118846).

\end{acknowledgments}

\begin{appendices}
\appendix
\section{Estimate of $\rho(S_n)$ 
and 
$\left<\Gamma_{\gamma 0} \right>$ for $^{64}$Zn }
\label{appA}
As briefly described in Sec.~\ref{sec:nldnorm}, we have used known $D_0$ and $D_1$ values for the other Zn isotopes to estimate the NLD at $S_n$ for $^{64}$Zn. 
We list all the input data we have used for this estimate in Table~\ref{tab:allNLDparametersEB} and Table~\ref{tab:allNLDparametersGC}. 
Moreover, the data used for the estimate of the total radiative widths are given in Table~\ref{tab:Gg}.

To convert the measured $D_0$ to the total level density, we insert $E_x=S_n$ into the expression for the spin distribution given by~\cite{Ericson59}
\begin{equation}
g(E_x,J) \simeq \frac{2J+1}{2\sigma_J^2(E_x)}\exp\left[-(J+1/2)^2/2\sigma_J^2(E_x)\right],
\label{eq:spindist}
\end{equation}
where $J$ is the spin quantum number. 
Further, we apply the functional form of the spin-cutoff parameter from Ref.~\cite{Capote09}:
\begin{equation}
\sigma_J^2(E_x)=\sigma_d^2 + \frac{\sigma_J^2(S_n)-\sigma_d^2}{S_n-E_d}\left(E_x-E_d\right),
\label{eq:sigE}
\end{equation}
where $\sigma_d$ is determined from known discrete levels at low excitation energy $E_x=E_d$, and $\sigma_J(S_n)$ is usually determined from global systematics.
In this case, we have applied two approaches to determine $\sigma(S_n)$, the empirical formula from Gilbert and Cameron (GC, Ref.~\cite{Gilbert1965}) and the more recent rigid-body moment of inertia expression from von Egidy and Bucurescu (EB, Ref.~\cite{egidy2005}).
As the shell-model calculations are in much better agreement with the EB approach, we have used that for the further analysis. 
For completeness, we include here the GC results also. 
All the $D_0$ and $D_1$ data, together with all the parameters used, can be found in Tables~\ref{tab:allNLDparametersEB} and ~\ref{tab:allNLDparametersGC} for the EB and GC spin cutoff parameters, respectively. 

To estimate the unknown $\rho(S_n, ^{64}\text{Zn})$, we assume the simplest possible model for the underlying trend of all the $\rho(S_n)$ values for the other Zn isotopes, namely an exponential increase in accordance with the constant-temperature model of Ericson~\cite{Ericson59}. 
For the linear regression we work on the natural-logarithmic scale, so that
\begin{equation}
    \ln \rho(S_n) \approx a + b S_n,
\end{equation}
where $a,b$ are constants to be determined from the fit. 

\begin{figure}[t]   \includegraphics[clip,width=1.\columnwidth]{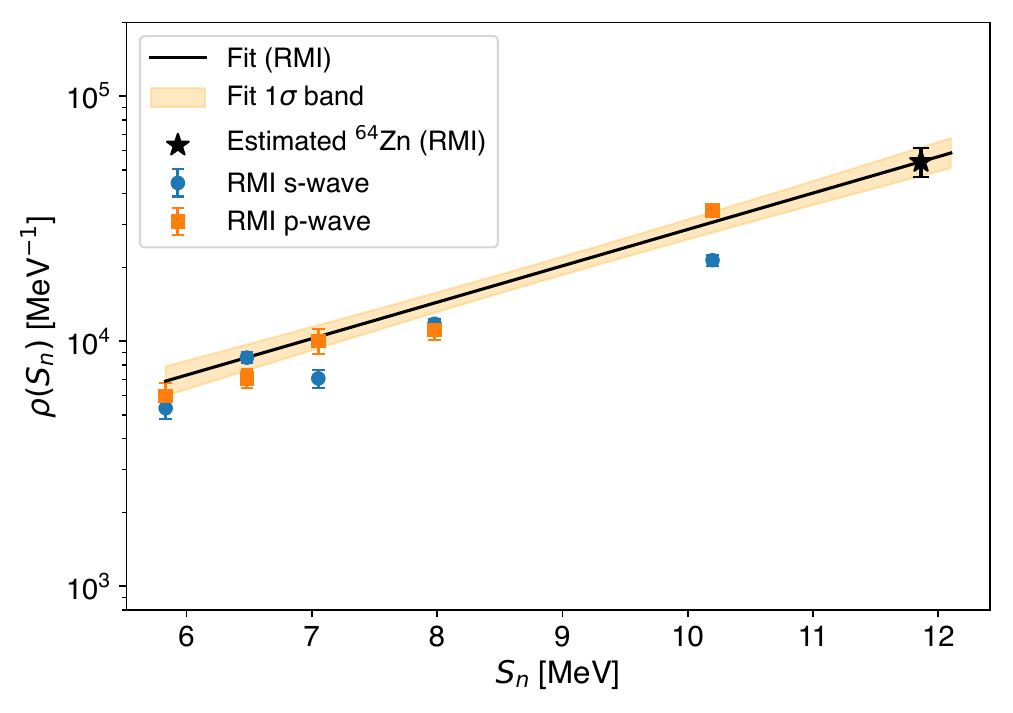}
    \caption{(Color online) Estimate of the level density $\rho(S_n)$ for $^{64}$Zn from a model fit to the semi-experimental $\rho(S_n)$ values for Zn isotopes where $s$-wave and $p$-wave resonance data exist, using the EB (RMI) spin-cutoff model.}
    \label{fig:rhoSn_est_EB}
\end{figure}
\begin{figure}[h]
    \includegraphics[clip,width=1.\columnwidth]{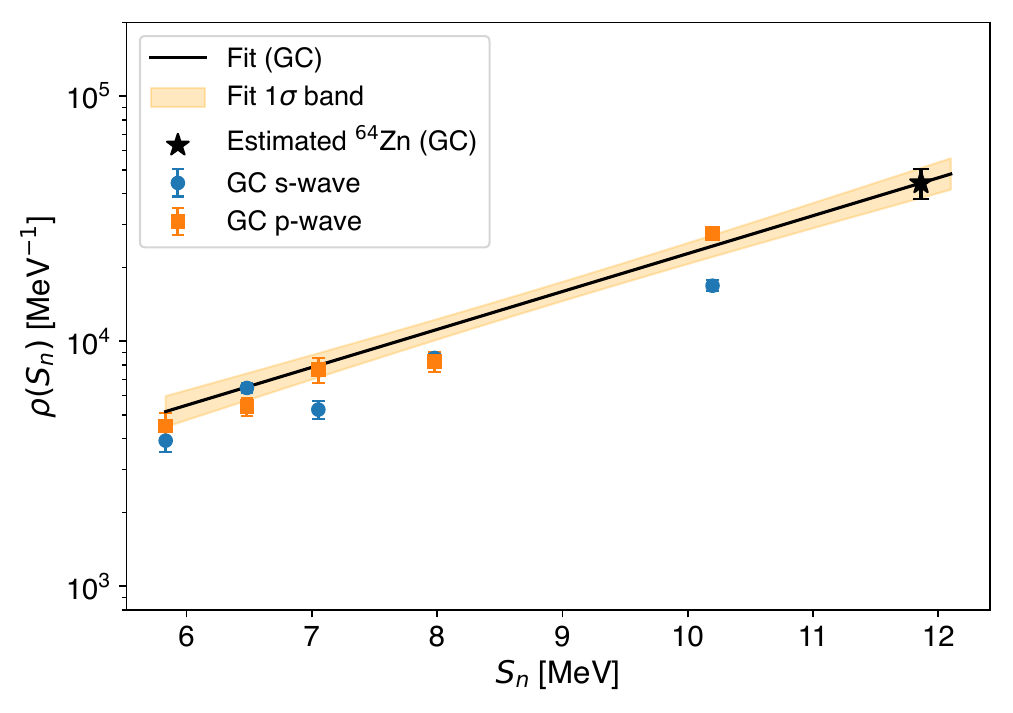}
    \caption{(Color online) Estimate of the level density $\rho(S_n)$ for $^{64}$Zn from a model fit to the semi-experimental $\rho(S_n)$ values for Zn isotopes where $s$-wave and $p$-wave resonance data exist, using the GC spin-cutoff model.}
    \label{fig:rhoSn_est_GC}
\end{figure}

As we use both the $s$-wave and the $p$-wave data, we introduce a mixed/hierarchical model.
This is because although the data do show a common trend,  which is an exponential increase with $S_n$, there are likely some systematic offsets in the two data sets. 
There are possible methodological problems to obtain a pure and complete data set of $s$-wave and $p$-wave neutron resonance data~\cite{Mughabghab2018}, and these issues could result in the calculated $s$-wave and $p$-wave $\rho(S_n)$ to not agree within their uncertainties, as seen in Tables~\ref{tab:allNLDparametersEB} and ~\ref{tab:allNLDparametersGC}.
To account for this, we apply the following hierarchical model:
\begin{equation}
    \ln \rho_{ij} = a + bS_{n,i} + \gamma_{\text{p}} I_{\text{wave=p}}(j) + \varepsilon_{ij},
\end{equation}
where $i$ indexes the isotopes, $j$ indexes the wave type ($s$ or $p$), $a$ is the intercept, $b$ is the slope, $\gamma_{\text{p}}$ is the offset of the $p$-wave data relative to the $s$-wave data (on the $\ln$ scale), 
$I_{\text{wave=p}}(j)$ is an indicator (1 for $p$‑wave, 0 for $s$‑wave), and $\varepsilon_{ij}$ are residual errors.
This is a basic two-level hierarchy, where level 1 is the underlying, method‑independent exponential trend (parameters $a,b$) and level 2 is the method effect $\gamma_{\text{p}}$ that captures the systematic shift between the $s$- and $p$-wave data.
\begin{figure}[h]
    \includegraphics[clip,width=1.0\columnwidth]{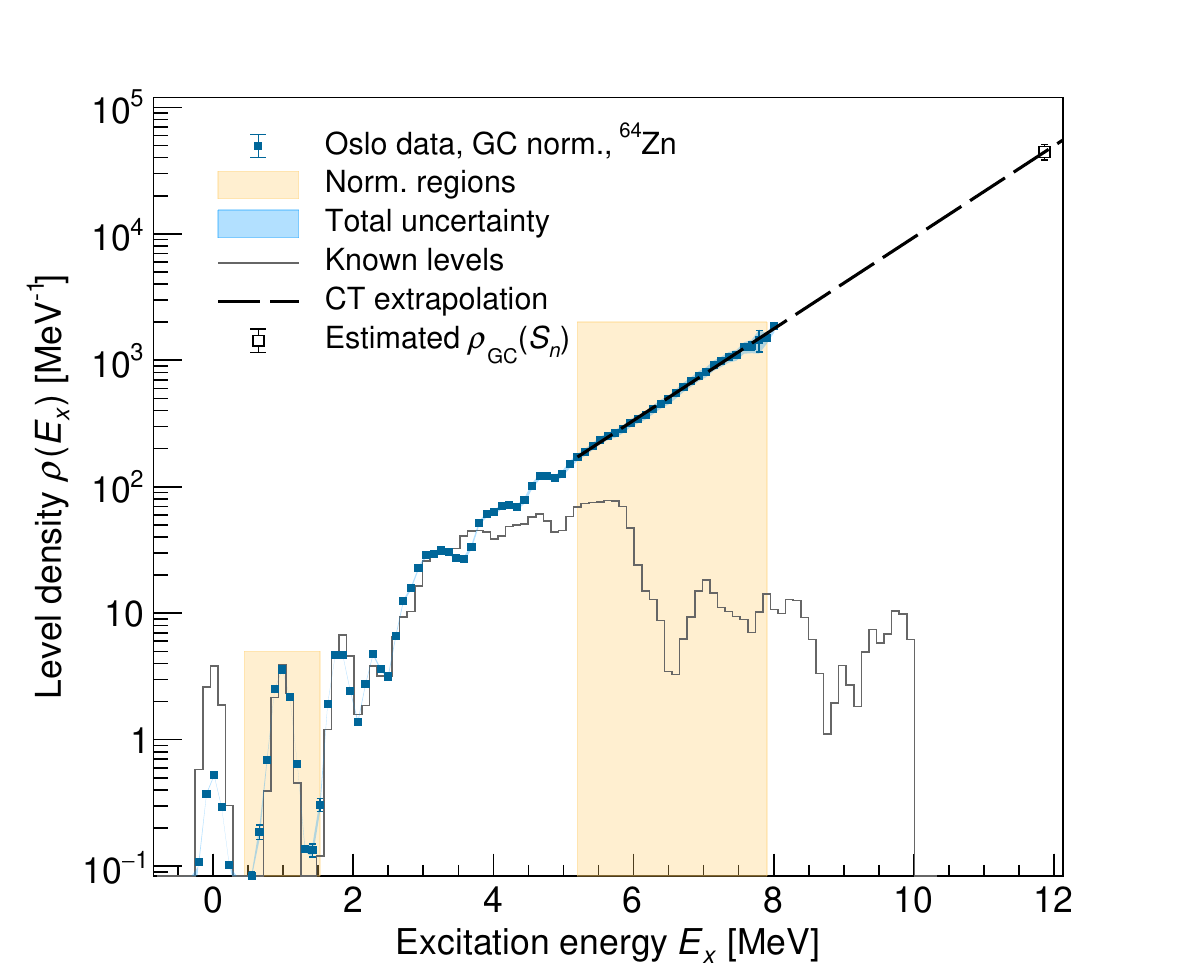}
    \caption{(Color online) Level density (filled data points) of $^{64}$Zn normalized to $\rho_{\mathrm{GC}}(S_n)$. 
    The data points are extrapolated by the dashed line to $\rho_{\mathrm{GC}}(S_n)$  using the CT model. 
    The histogram shows the level density of known levels~\cite{NNDC}, which has been smoothed with a Gaussian distribution with a FWHM of 230~keV.
    The shaded boxes indicate the regions used for fitting to discrete levels   and the CT model.
    The energy bins have a width of 108 keV.}
    \label{fig:counting_GC}
\end{figure}

Including the given uncertainties in the $s$ and $p$-wave data, we then perform a weighted least-squares fit through
\begin{equation}
 \ln \rho_{i,j} = \hat{a} + \hat{b}S_{n,i}  + \hat{\gamma}_{\text{p}} I_{\text{wave=p}}(j) + \varepsilon_{ij}
\end{equation}
to obtain the predicted value $\ln\rho(S_n, ^{64}\text{Zn})$, which is then transformed back to the original scale, and the uncertainty is obtained through a first-order error propagation to get the one-standard-deviation uncertainty (68.3\% confidence interval) on the linear scale: 
\begin{equation}
    \sigma_{\rho,{^{64}\text{Zn}}} \approx \rho(S_n, ^{64}\text{Zn})\sigma_{\ln \rho,{^{64}\text{Zn}}},
\end{equation}
where $\sigma_{\ln \rho,{^{64}\text{Zn}}}$ is obtained from the covariance matrix of the regression.

From the fit shown in Fig.~\ref{fig:rhoSn_est_EB}, we get $\rho_{\mathrm{EB}}(S_n,^{64}\mathrm{Zn}) = 5.41^{+0.78}_{-0.68}\times 10^4$ MeV$^{-1}$, and from the fit shown in  Fig.~\ref{fig:rhoSn_est_GC}, we find $\rho_{\mathrm{GC}}(S_n,^{64}\mathrm{Zn}) = 4.43^{+0.67}_{-0.58}\times 10^4$ MeV$^{-1}$.
Using the corresponding model for the spin-cutoff parameter of $^{64}$Zn at $S_n$, we obtain  $D_{0,EB} = 200^{+29}_{-25}$ eV  and $D_{0,GC} = 175^{+27}_{-23}$ eV.
The level density of $^{64}$Zn normalized to the GC value is shown in Fig.~\ref{fig:counting_GC}.

\begin{table*}[t]
\caption{Neutron-resonance data and input parameters used for the estimate of $\rho(S_n)$ of $^{64}$Zn. The target nucleus $A_t$ is the target in the ($n,\gamma$) reaction with $A$ nucleons, $S_n$ is the neutron separation energy of the $A+1$ compound nucleus, $J_t^{\pi}$ is the spin and parity of the ground state of the target nucleus, $D_0$ and $D_1$ are the $s$-wave and $p$-wave neutron resonance spacings from Ref.~\cite{Mughabghab2018},  $a_{\mathrm{EB}}$ is the level-density parameter for  von Egidy-Bucurescu global systematics, $E_{\mathrm{EB}}$ the corresponding energy shift,   $\sigma_{\mathrm{EB}}(S_n)$ the spin-cutoff parameter, $\rho_{\mathrm{EB,0}}(S_n)$ and $\rho_{\mathrm{EB,1}}(S_n)$ are the calculated level densities using the $D_0$ and $D_1$ level spacings, respectively.}
\begin{tabular}{cccccccccc}
\hline
\hline
$A_t$  & $S_n$  & $J_t^{\pi}$ & $D_0$ & $D_1$ & $a_{\mathrm{EB}}$ & $E_{\mathrm{EB}}$ & $\sigma_{\mathrm{EB}}(S_n)$ & $\rho_{\mathrm{EB,0}}(S_n)$ & $\rho_{\mathrm{EB,1}}(S_n)$ \\
          & [MeV] &       & [keV]    & [keV]      & [MeV$^{-1}$] & [MeV]  &         & [10$^3$ MeV$^{-1}$]  & [ 10$^3$ MeV$^{-1}$]      \\
\hline
$^{64}$Zn &7.97 & $0^{+}$& 2.94(13) & 1.10(10)   & 7.91 & -0.441 & 4.10 & 11.8(5) & 11.1(10)  \\
$^{66}$Zn &7.05 & $0^{+}$& 4.7(4)   & 1.17(10)   & 8.64 & -0.520 & 4.01 & 7.06(60)& 10.0(12)  \\
$^{67}$Zn &10.2 & $5/2^{-}$& 0.367(19) & 0.123(5)& 8.77 &  0.819 & 4.25 & 21.4(11)& 34.3(14)  \\
$^{68}$Zn &6.48 & $0^{+}$& 3.79(19) & 1.63(14)   & 9.17 & -0.483 & 3.97 & 5.43(47)& 7.07(61)  \\
$^{70}$Zn &5.83 & $0^{+}$& 6.00(60) & 1.90(24)   & 9.60 & -0.593 & 3.94 & 4.51(57)& 5.98(76)  \\

\hline
\hline
\end{tabular}
\\
\label{tab:allNLDparametersEB}
\end{table*}

\begin{table*}[t]
\caption{Same as Table~\ref{tab:allNLDparametersEB} but for the GC approach.  The target nucleus $A_t$ is the target in the ($n,\gamma$) reaction with $A$ nucleons, $S_n$ is the neutron separation energy of the $A+1$ compound nucleus, $J_t^{\pi}$ is the spin and parity of the ground state of the target nucleus, $D_0$ and $D_1$ are the $s$-wave and $p$-wave neutron resonance spacings from Ref.~\cite{Mughabghab2018}, $a_{\mathrm{GC}}$ is the level-density parameters for the Gilbert-Cameron global systematics, $E_{\mathrm{GC}}$ the corresponding energy shift,  $\sigma_{\mathrm{GC}}(S_n)$ the spin-cutoff parameter, $\rho_{GC,0}(S_n), \rho_{\mathrm{GC,1}}(S_n)$ are the calculated level densities using the $D_0$ and $D_1$ level spacings, respectively.}
\begin{tabular}{cccccccccc}
\hline
\hline
$A_t$  & $S_n$  & $J_t^{\pi}$ & $D_0$ & $D_1$ & $a_{\mathrm{GC}}$ & $E_{\mathrm{GC}}$ & $\sigma_{\mathrm{GC}}(S_n)$ & $\rho_{\mathrm{GC,0}}(S_n)$ & $\rho_{\mathrm{GC,1}}(S_n)$  \\
          & [MeV] &       & [keV]    & [keV]      & [MeV$^{-1}$] & [MeV]  &         & [10$^3$ MeV$^{-1}$]  & [10$^3$ MeV$^{-1}$]     \\
\hline
$^{64}$Zn &7.97 & $0^{+}$& 2.94(13) & 1.10(10)   &  7.93       & -0.923  & 3.47 & 8.55(38)& 8.26(75)  \\
$^{66}$Zn &7.05 & $0^{+}$& 4.7(4)   & 1.17(10)   &  8.15       & -0.992  & 3.44 & 5.26(45)& 7.65(92)  \\
$^{67}$Zn &10.2 & $5/2^{-}$& 0.367(19) & 0.123(5)&  8.25       &   1.19  & 3.57 & 16.9(9) & 27.5(11)  \\
$^{68}$Zn &6.48 & $0^{+}$& 3.79(19) & 1.63(14)   &  8.36       & -0.856  & 3.42 & 6.45(33)& 8.58(43) \\
$^{70}$Zn &5.83 & $0^{+}$& 6.00(60) & 1.90(24)   &  8.57       & -0.577  & 3.36 & 3.93(39)& 5.33(53) \\

\hline
\hline
\end{tabular}
\\
\label{tab:allNLDparametersGC}
\end{table*}

\begin{table}[t]
\caption{Neutron-resonance data and input parameters used for the estimate of  $\left< \Gamma_{\gamma 0}\right>$ of $^{64}$Zn. The target nucleus $A_t$ is the target in the ($n,\gamma$) reaction with $A$ nucleons, $S_n$ is the neutron separation energy of the $A+1$ compound nucleus, $J_t^{\pi}$ is the spin and parity of the ground state of the target nucleus, and $\left< \Gamma_{\gamma0} \right>$ is the average, total radiative width from $s$-wave neutron resonances taken from Ref.~\cite{Mughabghab2018}. }
\begin{tabular}{ccccc}
\hline
\hline
$A_t$  & $S_n$  & $J_t^{\pi}$ & $\left< \Gamma_{\gamma0} \right>$ Atlas 2018 & Used value\\
          & [MeV] &       &  [meV]  &  [meV]     \\
\hline
$^{64}$Zn &7.97 & $0^{+}$&  726(60) & 564(43) \\
$^{66}$Zn &7.05 & $0^{+}$&  400(20) & 365(29)\\
$^{67}$Zn &10.2 & $5/2^{-}$ & 440(60) & 429(18)\\
$^{68}$Zn &6.48 & $0^{+}$&  320(60) & 216(20)\\
$^{70}$Zn &5.83 & $0^{+}$&  198(30) & 198(30)\\

\hline
\hline
\end{tabular}
\\
\label{tab:Gg}
\end{table}

\begin{figure}[tb]
    \includegraphics[clip,width=1.\columnwidth]{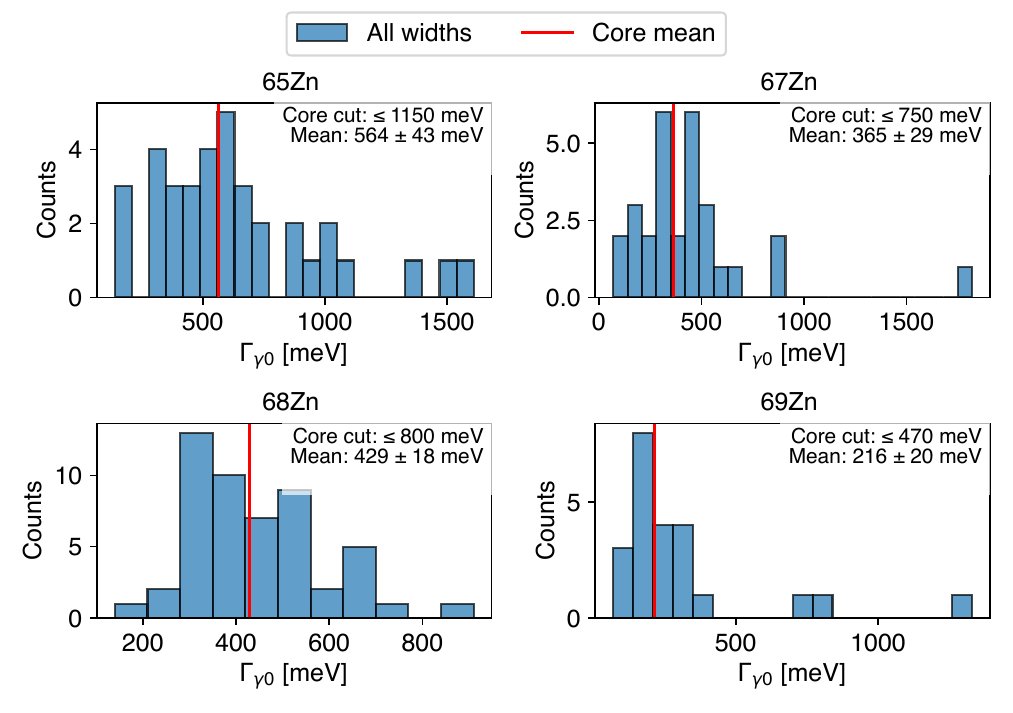}
    \caption{(Color online) Histograms of $ \Gamma_{\gamma0}$ values for Zn isotopes with 70 meV bin width. The red line shows the mean of the core-value distributions without outliers (core cut shown in the legends).}
    \label{fig:hist_Gammagamma}
\end{figure}

To obtain an absolute normalization of the $\gamma$SF, we have made use of available $s$-wave resonance data from Mughabghab's 2018 Atlas~\cite{Mughabghab2018}.
The  $\Gamma_{\gamma0}$ values for the  individual resonances of the compound nuclei $^{65,67,68,69}$Zn are binned and shown as histograms in Fig.~\ref{fig:hist_Gammagamma}.
As can be seen from Fig.~\ref{fig:hist_Gammagamma}, the distributions of $\Gamma_{\gamma0}$ values are quite broad and skewed, with a few very large widths that sit far out in the high-value tails.
Due to the presence of these outliers, we have made cuts for each data set (as indicated in the legends) before re-calculating the average, total radiative widths $\left<\Gamma_{\gamma0}\right>$. 
These outliers could perhaps be due to resonances with a doorway-like structure leading to an increased $\gamma$-decay width.
The new estimated uncertainties for the $\left<\Gamma_{\gamma0}\right>$ are then based on the sample spread, i.e. the standard deviation across individual widths after removing the outliers, rather than the quoted uncertainties for each individual resonance. 
Specifically, we calculate the standard error \textit{SE} from the standard deviation \textit{SD} of the core mean as 
\begin{equation}
    SE = \frac{SD}{\sqrt{N_{\text{core}}}},
\end{equation}
where $N_{\text{core}}$ is the number of core values.
The re-estimated values with uncertainties (standard errors) are listed in Table~\ref{tab:Gg}.
Note that $^{71}$Zn (target nucleus $A_t$ is $^{70}$Zn) has no firm $\Gamma_{\gamma0}$ values for any of the $s$-wave resonances, we therefore used Mughabghab's estimate for this case.

\begin{figure}[t]
    \includegraphics[clip,width=1.0\columnwidth]{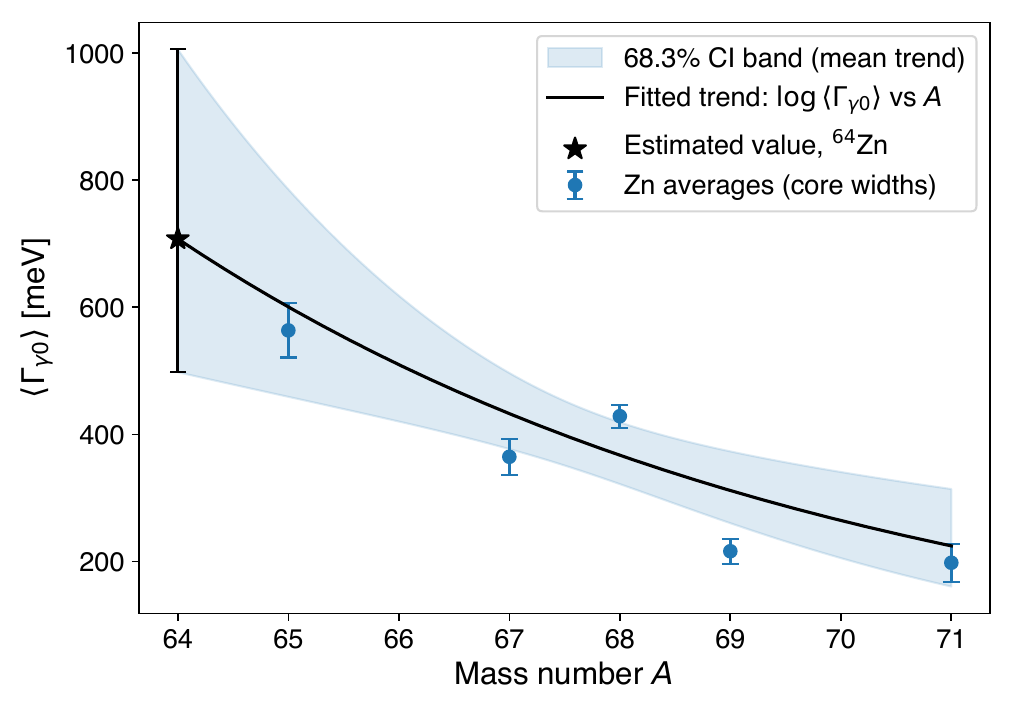}
    \caption{(Color online) Estimate of $\left< \Gamma_{\gamma0}\right>$ for $^{64}$Zn (see text).}
    \label{fig:Gammagamma}
\end{figure}

\begin{figure}[t]
    \includegraphics[clip,width=1.0\columnwidth]{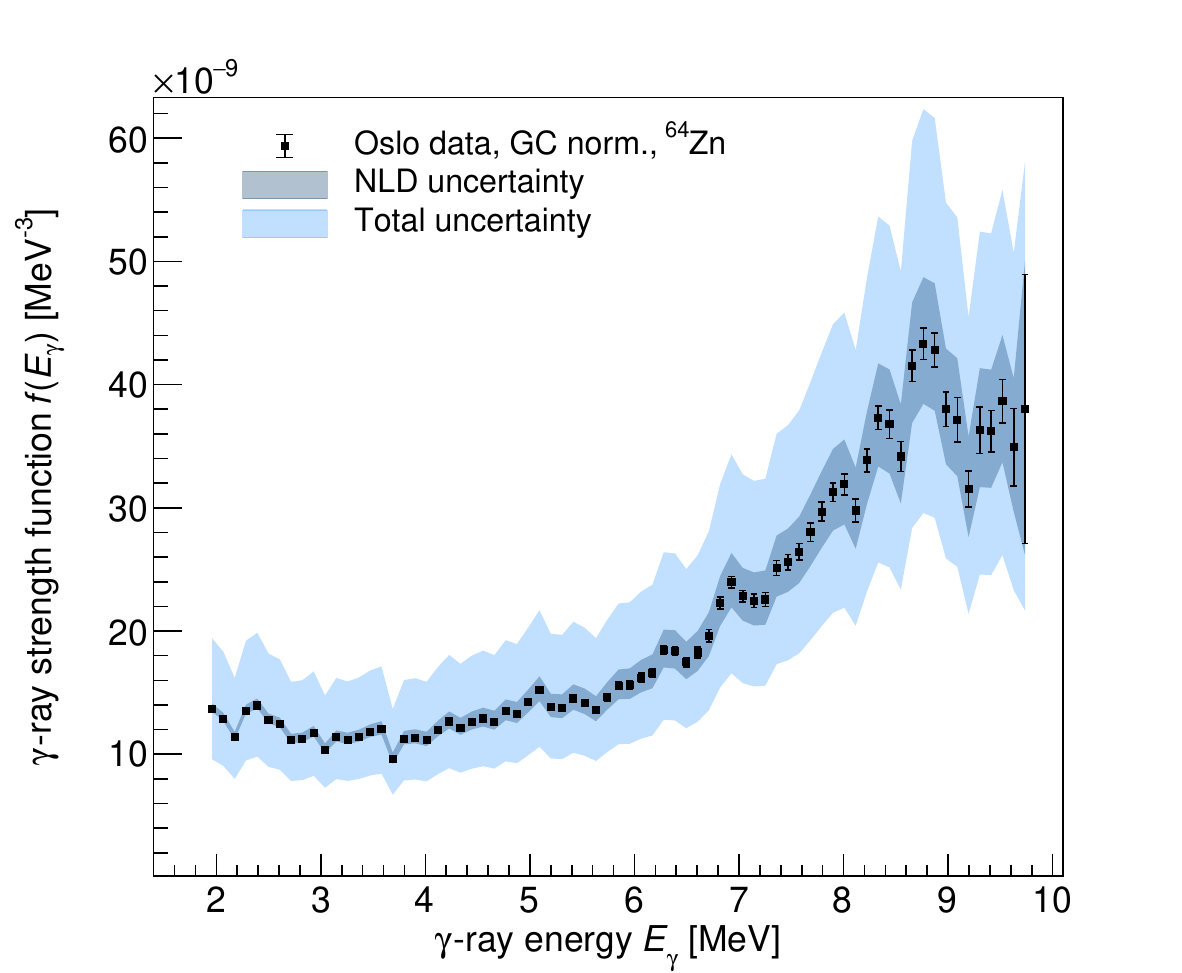}
    \caption{(Color online) Normalized $\gamma$SF of $^{64}$Zn for the GC normalization (see text). The energy bins have a width of 108 keV.}
    \label{fig:gsfGC}
\end{figure}

We now apply a simple model to fit the $\left<\Gamma_{\gamma0}\right>$ values and estimate the unknown value for $^{64}$Zn:
\begin{equation}
 \ln \left< \Gamma_{\gamma0} \right> = c + d A,
\end{equation}
where $A$ is the mass number of the compound nucleus after neutron capture.
The result of the regression (weighted least squares) with the fit is shown in Fig.~\ref{fig:Gammagamma}, 
where the $1\sigma$ confidence band (transformed back to linear scale) is shown together with the estimated value for $^{64}$Zn, $\left< \Gamma_{\gamma0}(^{64}\text{Zn}) \right> =  708^{+298}_{-210}$ meV. 
As seen in the figure, the trend seems to be rather well captured by the simple model.
For completeness, we also show the resulting normalized $\gamma$SF for the GC approach in Fig.~\ref{fig:gsfGC}.

\section{Hindered transitions to the ground state}
\label{appB}
The experimental NLD extracted with the Oslo method  depends on the intensity for which final levels $J_f$ are populated from higher lying initial levels $J_i$. 
This intensity depends on many different $\gamma$-ray energies as the range of initial excitation energies $E_i$ is large. 
Considering only dipole transitions, all final levels with $J_f>0$ are populated by six types of initial levels $J_i$: $J_i=J_f-1$, $J_i=J_f$ and $J_i=J_f+1$ and both parities.
In contrast, levels with $J_f=0$  are populated by only two types of initial levels, namely $J_i=1^+$ and $1^-$. 
In the Oslo method, these spin-selection rules lead to an underestimate of the NLD of $J=0$ levels, as discussed in Sec.~\ref{sec:shapemethod}. 
However, the fact that we find as little as $N\approx0.13$ for the $0^+$ ground level is surprising. 
\begin{figure}[t]
    \includegraphics[clip,width=0.9\columnwidth]{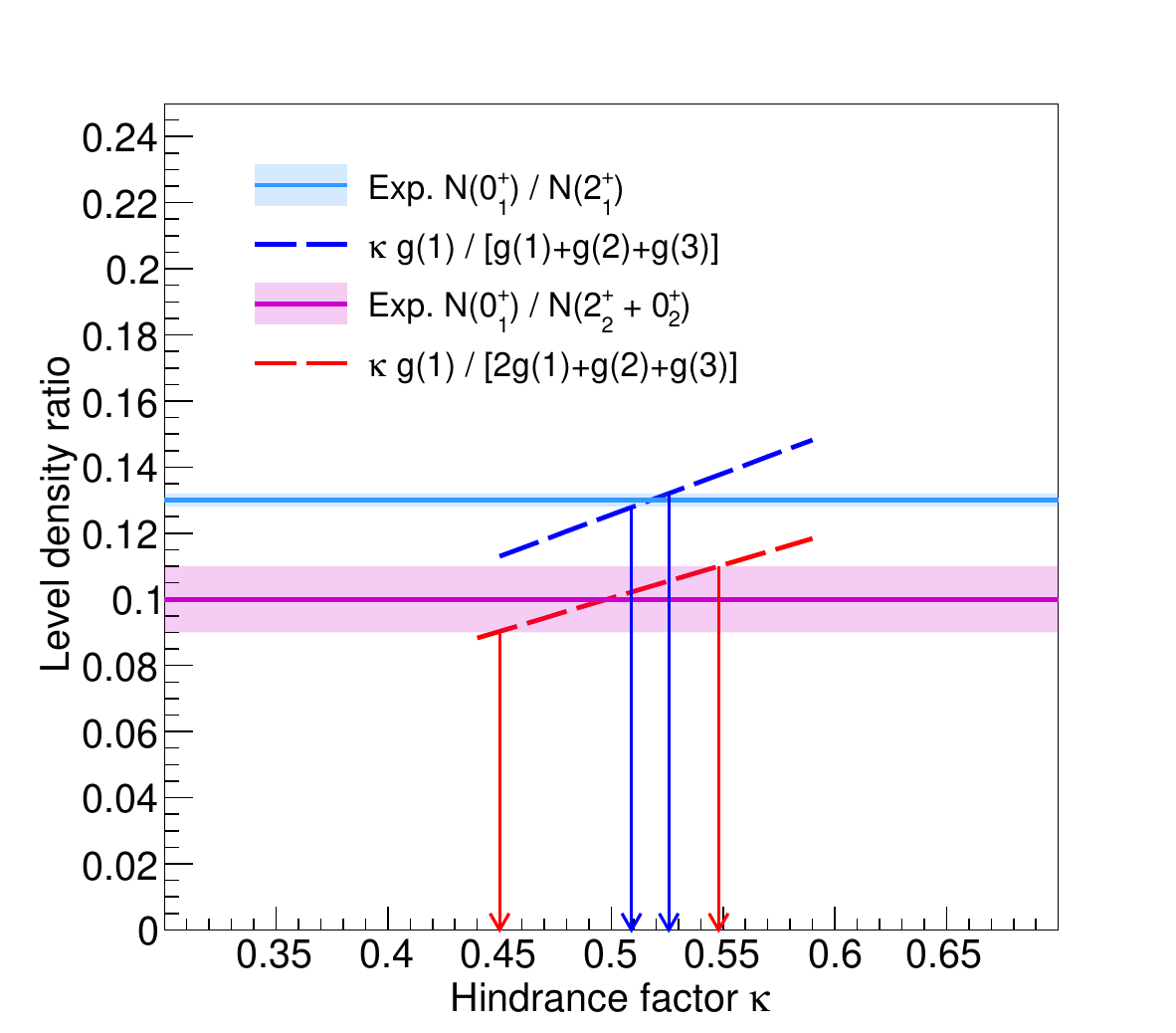}
    \caption{(Color online) Experimental level-density ratios (horizontal lines with error band) obtained from the number of levels $N$ found from the integrated level density of the levels $0^+_\text{gs}$, $2^+_1$, and $2^+_2,0^+_2$. The ratios are compared with the fraction of the spin population feeding the levels of the three bumps according to hypothesis H1.}
    \label{fig:zerotwox}
\end{figure}

We have tested three hypotheses (hypothesis H1, H2, and H3) to arrive at a consistent explanation of the three numbers of levels $N$ displayed in Fig.~\ref{fig:countingx_lin}. 
Here, we use the average initial excitation energy  for the range $E_i = 4.98-9.84$ MeV giving $\left<E_i\right> = 7.41$ MeV, with the corresponding spin-cutoff parameter from Eq.~(\ref{eq:sigE}) $\sigma_J = 3.53$. 

This spin-cutoff parameter is then used in the spin distribution $g(\left<E_i\right>,J)$ to estimate the expected ratios of the number of levels for $N(0_\text{gs}^+)/N(2_1^+)$, $N(0_\text{gs}^+)/N(2_2^+ +0^+_2)$ and  $N(2_1^+)/N(2_2^+ +0_2^+)$.
As we operate with the average, initial excitation energy we omit this in the notation for $g$ in the following. 
From integration of the Oslo NLD data, we have $N(0_\text{gs}^+)/N(2_1^+)\approx 0.13$, $N(0_\text{gs}^+)/N(2_2^+ + 0_2^+)\approx 0.10$ and $N(2_1^+)/N(2_2^+ + 0_2^+) \approx 0.77$.
These are the numbers we compare the following hypotheses to:

\noindent(H1) \textit{Transitions to the ground state are hindered}.\\
If the transitions to the ground state are hindered with a hindrance factor $\kappa$, then this results in the following ratios:
\begin{eqnarray}
\frac{N(0_\text{gs}^+)}{N(2_1^+)}&=&\frac{\kappa g(1)}{g(1)+g(2)+g(3)}
\label{eq:array1_1}\\
 \frac{N(0_\text{gs}^+)}{N(2_2^+ + 0_2^+)}&=&\frac{\kappa g(1)}{2g(1)+g(2)+g(3)}
\label{eq:array2_1}\\
 \frac{N(2_1^+)}{N(2_2^+ + 0_2^+)}&=&\frac{g(1)+g(2)+g(3)}{2g(1)+g(2)+g(3)}
\label{eq:array3_1}
\end{eqnarray}

\noindent(H2) \textit{Transitions to both $0^+$ levels are hindered}.\\
If there is a common hindrance factor $\kappa$ for both the ground state and the  excited $0^+_2$, then we obtain:
\begin{eqnarray}
\frac{N(0_\text{gs}^+)}{N(2_1^+)}&=&\frac{\kappa g(1)}{g(1)+g(2)+g(3)}
\label{eq:array1_2}\\
 \frac{N(0_\text{gs}^+)}{N(2_2^+ + 0_2^+)}&=&\frac{\kappa g(1)}{g(1)+g(2)+g(3)+\kappa g(1)}
\label{eq:array2_2}\\
 \frac{N(2_1^+)}{N(2_2^+ + 0_2^+)}&=&\frac{g(1)+g(2)+g(3)}{g(1)+g(2)+g(3)+\kappa g(1)}
\label{eq:array3_2}
\end{eqnarray}

\noindent(H3) \textit{The spin distribution is too large for $J=1$}.\\
Instead of a hindrance factor for either the ground state or both  $0^+$ levels, it could be that the assumed smooth spin distribution has overestimated the real $J=1$ fraction. This would result in the $\kappa$ factor not being a hindrance factor, but a correction to the too large $J=1$ level density. 
If this would be the case, we have the following relations:
\begin{eqnarray}
\frac{N(0_\text{gs}^+)}{N(2_1^+)}&=&\frac{\kappa g(1)}{\kappa g(1)+g(2)+g(3)}
\label{eq:array1_3}\\
 \frac{N(0_\text{gs}^+)}{N(2_2^+ + 0_2^+)}&=&\frac{\kappa g(1)}{2\kappa g(1)+g(2)+g(3)}
\label{eq:array2_3}\\
 \frac{N(2_\text{1}^+)}{N(2_2^+ + 0_2^+)}&=&\frac{\kappa g(1)+g(2)+g(3)}{2\kappa g(1)+g(2)+g(3)}
\label{eq:array3_3}
\end{eqnarray}
Figure~\ref{fig:zerotwox} shows that Eqs.(\ref{eq:array1_1}) and (\ref{eq:array2_1}) are fulfilled within the experimental errors for a hindrance factor of $\kappa=0.52(1)$ and $\kappa=0.50(5)$, respectively. 
For the shape method in Sec.~\ref{sec:shapemethod} we have applied $\kappa=0.52$, in accordance with these estimates. 
The estimated ratios from Eq.~(\ref{eq:array1_1}) and Eq.~(\ref{eq:array2_1}) are 0.131 and 0.104, which agree very well with the experimental values 0.130(1) and 0.10(1).
The estimated ratio based on the $g(J)$ values of Eq.~(\ref{eq:array3_1}), which is independent on $\kappa$, gives a constant 0.799 that is fully consistent with the experimental value of $0.77(9)$. 
Hypothesis H2, for a ratio ${N(0_\text{gs}^+)}/{N(2_1^+)} = 0.131$ in accordance with the experimental one, gives ratios ${N(0_\text{gs}^+)}/{N(2_2^+ + 0_2^+)} = 0.116$ and ${N(2_\text{1}^+)}/{N(2_2^++ 0_2^+)}= 0.885$.  
On the other hand, for ${N(0_\text{gs}^+)}/{N(2_1^+)} = 0.131$, H3 leads to ratios  ${N(0_\text{gs}^+)}/{N(2_2^++0_2^+)} = 0.116$ and ${N(2_1^+)}/{N(2_2^+ +0_2^+)}= 0.884$, and this implies a suppression of $J=1$ levels with a factor $\kappa = 0.45$.
We conclude that the hypothesis that agrees best with the data is H1, although hypotheses H2 and H3 cannot be completely ruled out. 

\section{Additional experimental details, quality checks and convergence results for the Oslo-method steps}
\label{appC}
The $^{64}$Zn experiment at OCL ran from November 19 to November 23, 2018, with the full OSCAR array and with digital  electronics. The raw matrix was generated with a gate on the prompt time peak for the time difference between the particle and $\gamma$ detection, and contained 3.94$\times 10^8$ counts. 
The background matrix generated with a gate on a time-spectrum peak for a later beam pulse had 1.80$\times10^7$ counts. i.e. less than 5\% of the prompt-gate counts. 

The unfolding of the $\gamma$-ray spectra was performed using the iterative method presented in Ref.~\cite{Gut96}, with a goodness-of-fit controlled through a 20\% weight on the experimental fluctuations (calculated from an average of neighboring bins according to the resolution in that $\gamma$ energy region) and an 80\% weight on the $\chi^2$ minimization of the comparison with the raw and folded spectrum. 
Typically, with a bin width of 36 keV, about 100 iterations are sufficient to reach convergence. 

\begin{figure*}[t]
\includegraphics[clip,width=1.8\columnwidth]{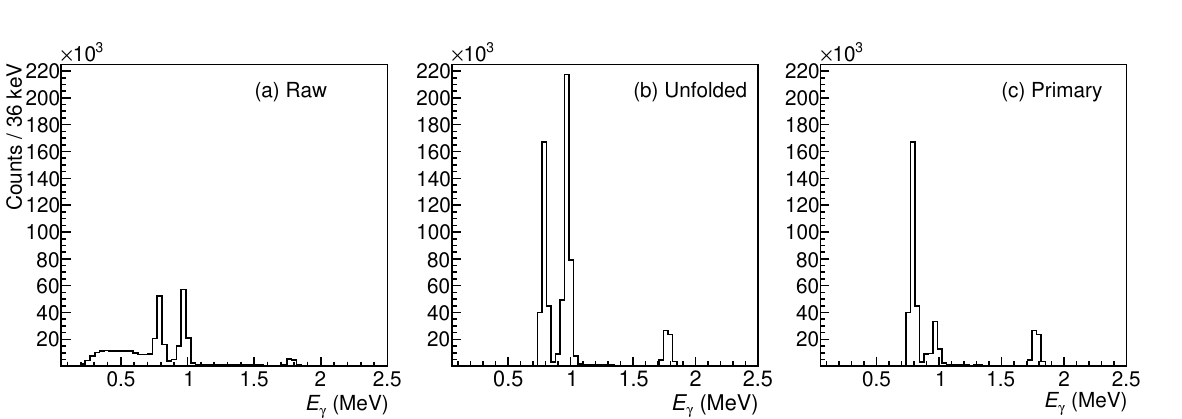}
\caption{The raw (a), unfolded (b) and primary (c) $\gamma$-ray spectrum gated on  $E_x \approx 1.80$ MeV, including the $2_2^+$ and  $0_2^+$ levels in $^{64}$Zn. 
}
\label{fig:gammaspectra}
\end{figure*}

\begin{figure*}[t]
\includegraphics[clip,width=1.8\columnwidth]{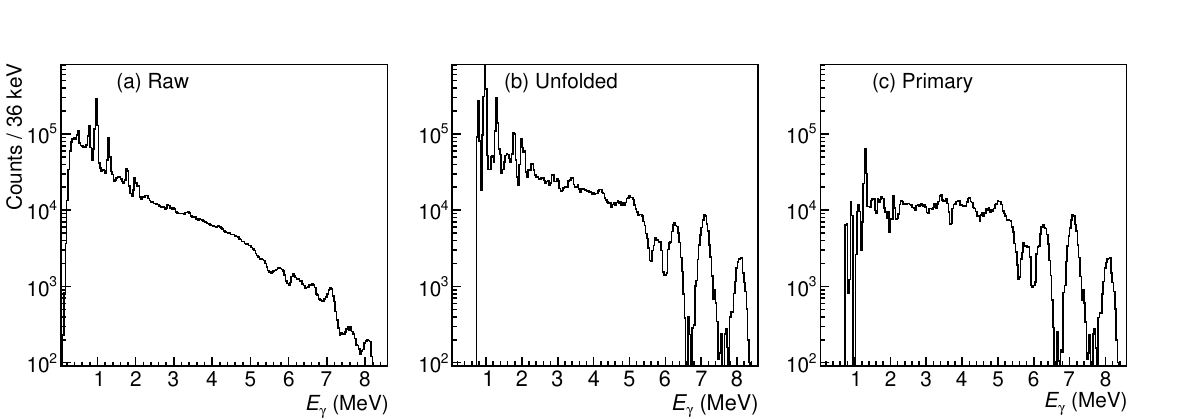}
\caption{The raw (a), unfolded (b) and primary (c) $\gamma$-ray spectrum gated on  $E_x \approx 8.0$ MeV in $^{64}$Zn. 
}
\label{fig:gammaspectra_highEx}
\end{figure*}

The first-generation method makes use of an area-consistency check for the obtained primary spectra to ensure that the $\gamma$-ray multiplicity for the primary spectrum at a given $E_i$ bin is $\approx 1$. 
The allowed correction factor $c_\text{corr}$ to the weighting coefficients $w_{E_i}$ does not exceed 15\%, i.e. $c_\text{corr} \in [0.85,1.15]$, and a $c_\text{corr}$ value close to unity indicates that the correct weighting coefficient (and hence the primary spectrum) was found for this specific $E_i$.
Within the excitation-energy range used for the extraction of the NLD and $\gamma$SF, i.e. $6.0 \text{ MeV} <E_i < 10 \text{ MeV}$, $c_\text{corr} \in [1.00,1.03]$, indicating full convergence and no significant adjustment to the weighting coefficients.   
An example of the raw, unfolded and first-generation spectrum with a gate on $E_x = 1.80$ MeV is shown in Fig.~\ref{fig:gammaspectra}, and with a gate on $E_x = 8.0$ MeV in Fig.~\ref{fig:gammaspectra_highEx}.
We applied the requirement that $E_\gamma > 2.0$ MeV for the NLD and $\gamma$SF extraction, avoiding the low-$E_\gamma$ region where some non-primary transitions might still be present.

\end{appendices}

\end{document}